\documentclass[sigconf]{acmart}

\usepackage{multirow}
\usepackage{graphicx}
\usepackage[export]{adjustbox}
\usepackage{subcaption}
\usepackage{lipsum}
\usepackage{algorithm}
\usepackage{algpseudocode}
\usepackage{amsmath}
\usepackage{xspace}

\AtBeginDocument{%
  }

\setcopyright{acmlicensed}
\copyrightyear{2018}
\acmYear{2018}
\acmDOI{XXXXXXX.XXXXXXX}
\acmConference[Conference acronym 'XX]{Make sure to enter the correct
  conference title from your rights confirmation email}{June 03--05,
  2018}{Woodstock, NY}
\acmISBN{978-1-4503-XXXX-X/2018/06}

\ifodd 0
\definecolor{darkgreen}{RGB}{0,200,0}
\newcommand{\newrev}[1]{{\color{red}#1}} 
\newcommand{\yuxinzhangrev}[1]{{\color{purple}#1}}
\newcommand{\lizhrev}[1]{{\color{blue}#1}}

\else
\newcommand{\yuxinzhangrev}[1]{#1} 
\newcommand{\lizhrev}[1]{#1} 

\newcommand{\newrev}[1]{#1} 
\fi

\newcommand{\name}{Grace\xspace}

\begin{document}

\title{Enabling Near-realtime Remote Sensing via Satellite–Ground Collaboration of Large Vision–Language Models}


\author{Zihan Li}
\affiliation{%
  \institution{Fudan University}
  \city{Shanghai}
  \country{China}}

\author{Jiahao Yang}
\affiliation{%
  \institution{Fudan University}
  \city{Shanghai}
  \country{China}}

\author{Yuxin Zhang}
\affiliation{%
 \institution{Fudan University}
  \city{Shanghai}
  \country{China}}

\author{Zhe Chen}
\affiliation{%
 \institution{Fudan University}
  \city{Shanghai}
  \country{China}}


\author{Yue Gao}
\affiliation{%
 \institution{Fudan University}
  \city{Shanghai}
  \country{China}}



\begin{abstract}
\yuxinzhangrev{Large vision–language models~(LVLMs) have recently demonstrated great potential in remote sensing (RS) tasks (e.g., disaster monitoring) conducted by low Earth orbit (LEO) satellites.
However, their deployment in real-world LEO satellite systems remains largely unexplored, hindered by limited onboard \newrev{computing} resources and \newrev{brief satellite–ground contacts}.}
\yuxinzhangrev{We propose \name, a satellite–ground collaborative system designed for near-realtime LVLM inference in RS tasks.}
\newrev{Accordingly, we deploy compact LVLMs on satellites for realtime inference, but larger ones on ground stations~(GSs) to guarantee end-to-end performance. \name is comprised of two main phases that are asynchronous satellite-GS Retrieval-Augmented Generation~(RAG), and a task dispatch algorithm. Firstly, we distill the knowledge archive of GS RAG to satellite archive with tailored adaptive update algorithm during limited satellite-ground data exchange period. Secondly, propose a confidence-based test algorithm that either processes the task onboard the satellite or offloads it to the GS.}
Extensive experiments \lizhrev{based on real-world satellite orbital data} show that \name reduces \lizhrev{the average latency by 76–95\%} compared to state-of-the-art methods, without compromising inference accuracy.

\end{abstract}




\maketitle

\section{Introduction}

\begin{figure}[t]
    \centering
    \includegraphics[width=1\columnwidth]{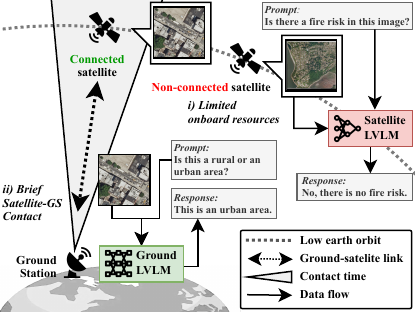}
    \caption{\lizhrev{Though collaborative inference may facilitate an efficient LVLM inference on satellites, i) limited onboard resources and ii) brief satellite-GS contact still pose significant challenges to the deployment.}}
    \Description{A diagram illustrating collaborative inference on satellites, highlighting limited onboard resources and brief satellite-ground station contact as challenges.}
    \label{fig:intro}
\end{figure}

In recent years, advances in satellite technology and declining launch costs have \yuxinzhangrev{driven a rapid proliferation of Low Earth Orbit (LEO) satellites for Remote Sensing (RS)}~\cite{zhang2025satellite,cheng2025geoduck, singh2024spectrumize,lin2024fedsn,zhao2024leo,yuan2024satsense,peng2025sigchord,lin2025esl}.
For example,
\lizhrev{Planet has launched 452 satellites to construct the Dove constellation, imaging more than 350 million square kilometer every day, delivering the near-daily, high-frequency coverage \cite{planet,yuan2025constructing,peng2024sums,lin2025leo,yuan2023graph}.}
\lizhrev{Positioned at altitudes of up to 2000 kilometers and equipped with advanced sensors, LEO satellites take full advantage of their unique vantage points to acquire wide-swath and high-resolution images of the Earth’s surface~\cite{zhang2024satfed}.}
\lizhrev{Leveraging AI model for information analysis, the vast volume of satellite data can be processed efficiently, \newrev{especially}, latency-sensitive applications \newrev{such as} forest fire detection \cite{barmpoutis2020review}, maritime surveillance \cite{do2023skysea}, and urban traffic monitoring \cite{vaidya2024exploiting, gersey2025sniffing, adkins2018signpost, yan2025large}}. 

\lizhrev{However, current latency-sensitive RS tasks primarily rely on traditional small models (e.g., CNN \cite{kattenborn2021review} and ViT \cite{wang2022advancing}), whose limitations are becoming increasingly evident:}
1) Small models require task-specific architectural design and parameter training, significantly limiting their adaptability across diverse RS applications. 
2) With only tens of millions of parameters, they struggle to capture the complex features inherent in satellite data, resulting in suboptimal performance \cite{kaplan2020scaling, xu2022simple}.
Fortunately, recent breakthroughs in Large Vision-Language Models (LVLMs) have overcome these limitations, revealing transformative potential for RS tasks \cite{zhang2024earthgpt, ouyang2024llmsense}.
By integrating visual and language understanding within a unified architecture, LVLMs achieve a ``one-model-for-all-tasks'' paradigm, thereby offering broad adaptability \cite{hu2024large, wang2024multimodal, sarhaddi2025llms, hota2024evaluating, ji2024mindguard}. Besides, with typically over a billion parameters, LVLMs show strong capabilities in fitting and generalizing for satellite data, significantly enhancing RS performance \cite{li2024vision}.
\textbf{Challenges.}
\yuxinzhangrev{Despite their advantages, the heavy computational demands of LVLMs present a core challenge: \textit{How to deploy them efficiently within the strict resource limits of LEO satellite systems}, constrained by two fundamental factors.}
\lizhrev{
\textit{First,} LEO satellites must prioritize onboard payload and energy for control systems, remote image acquisition, communicating with GSs, and other critical functions. This allocation leaves insufficient capacity to support large-scale LVLMs \cite{li2024dual, li2023boosting, li2024cwgan,chen2024gradient,zhang2025lcfed}, resulting in critically constrained computational resources. While modern LEO satellites are equipped with some edge computing platforms (like NVIDIA Jetson) to enable AI workloads \cite{satellitejetson}, these embedded systems have significant gaps in computing capability and memory resources compared to ground-based server clusters --- a significant limitation given the resource demands of LVLMs \cite{sooriya2023poster,cai2024self,hu2025lightllm}.
\textit{Second,} the unique network conditions of LEO satellites, which move at the orbital velocity that results in highly intermittent communication windows with ground stations \cite{starlinkorb}. During these long periods of disconnection, satellites must rely solely on their limited onboard processing capabilities for real-time data handling \cite{zhang2025s,lin2024fedsn, zhai2023fedleo,lin2025esl}.
The satellite-ground communication bandwidth also presents a critical bottleneck. Due to the inherent temporal constraints of satellite-ground links, satellites accumulate substantial volumes of remote sensing imagery during disconnection periods \cite{shenoy2024cosmac, singh2024exploiting,zhang2024satfed}. For high-resolution Earth observation systems, this results in persistent data backlogs that cannot be fully transmitted during brief communication windows. The resulting constrained data transmission capacity limits the system's ability to perform timely processing and analysis \cite{vasisht2021l2d2}.}

Existing LVLM deployments \yuxinzhangrev{in satellite systems} fall short of addressing these challenges and can be broadly categorized into two types:
1) Ground station-centric deployment schemes \cite{kuckreja2024geochat, zhang2024earthgpt}, while free from onboard computational limitations and able to leverage the powerful computing resources on the ground, are constrained by the bandwidth limitation and intermittent connectivity, making it ineffective to transmit large volumes of satellite data efficiently. The latency from data acquisition to analysis results is prolonged, failing to meet the demands for efficient inference;
2) Conversely, onboard deployment schemes \cite{sarhaddi2025llms, sooriya2023poster}, although reducing dependency on network bandwidth, are limited by the scarce computational resources on satellites, preventing the execution of large-scale LVLMs and thereby restricting intelligent inference capabilities. 
\lizhrev{Collaborative inference \cite{wang2023tabi}, a paradigm that partitions tasks between lightweight models on edge devices and powerful cloud models to optimize latency and resource efficiency, faces challenges in satellite environments. Existing implementations require sustained network connections, which are incompatible with the intermittent connectivity characteristics of LEO satellites.}

\textbf{Solution.}
\yuxinzhangrev{To tackle the above challenges,
we propose \name as a collaborative LEO satellite-GS LVLMs inference system for near-realtime RS.
\name deploys compact LVLMs onboard satellites for real-time inference, while larger models on the ground provide auxiliary support during each contact window.}
\lizhrev{To overcome the first challenge, we need to improve the accuracy of a compact LVLM. A common approach is to construct a external knowledge archive with Retrieval-Augmented Generation (RAG) to equip the model with domain-specific knowledge and reduce hallucinations. However, due to challenges mentioned earlier, it is necessary to minimize the size of the on-satellite RAG system to fit within the limited onboard resources. When RS missions change, efficiently updating the content in the satellite archive to adapt to new tasks becomes an issue need to be addressed. Therefore, we propose a dynamic satellite-ground collaborative RAG system. Similar to the collaborative inference system, this RAG system leverages a comprehensive ground archive and a streamlined satellite archive to enable efficient LVLM inference. To ensure that the satellite archive remains aligned with current mission requirements, we propose an adaptive update algorithm. This algorithm enables data exchange between satellites and GSs during their brief communication window. The GS customizes and updates the content of the satellite archive based on queries that are difficult for the satellite to handle, ensuring that the archive meets the latest mission needs.}
\lizhrev{To address the second challenge, we design a task dispatcher to reduce data transmission requirements. The task dispatcher strategically assigns inference tasks. With the confidence-based test, most tasks are processed onboard, significantly reducing the number of samples that must be transmitted to the GS. The dispatcher first need to determine whether the prior knowledge in the satellite archive is sufficient to support answering the query. Next, the dispatcher assess the confidence level of the satellite LVLM's prediction. If the prior knowledge is inadequate or the confidence in the prediction is low, the onboard inference result should be discarded, and the query must be transmitted to the GS for further processing. Otherwise, the onboard inference result is accepted as the final output.}

Grace maintains efficient data processing capabilities and inference performance through these measures with constrained computational resources and unstable network connectivity. In summary, this work has the following contributions:

 
\begin{itemize}
    \item \lizhrev{To the best of our knowledge, Grace --- a collaborative LEO satellite-GS inference system --- represents the first work of near-realtime LVLM inference in LEO satellite networks.}
    \item \lizhrev{Grace effectively addresses the onboard computational constraints of LEO satellites through its implementation of a dynamic knowledge archive, which enables adaptive content updates.}
    \item We develop and integrate key mechanisms, including the task dispatcher, to optimize data handling and enhance communication efficiency under constrained resources.
    \item Extensive evaluations demonstrate that our framework substantially outperforms other deployment baselines. \lizhrev{Grace reduces the average latency by 76–95\% compared to state-of-the-art methods, without compromising inference accuracy.}
\end{itemize}

The structure of the paper is as follows. In Section \ref{sec:challenges-and-motivation}, we highlight the challenges faced by current LEO satellite networks. Section \ref{sec:system-design} outlines the system design of Grace. Section \ref{sec:implementation} describes the implementation of the system, while Section \ref{sec:performance-evaluation} provides an evaluation of its performance. A discussion on related works is presented in Section \ref{sec:related-work-and-discussion}. The paper concludes with Section \ref{sec:conclusion}.

\section{Challenges and Motivation}
\label{sec:challenges-and-motivation}

\begin{figure*}[t]
    \centering
    \begin{subfigure}[b]{0.245\linewidth}
        \centering
        \includegraphics[width=\columnwidth, valign=c]{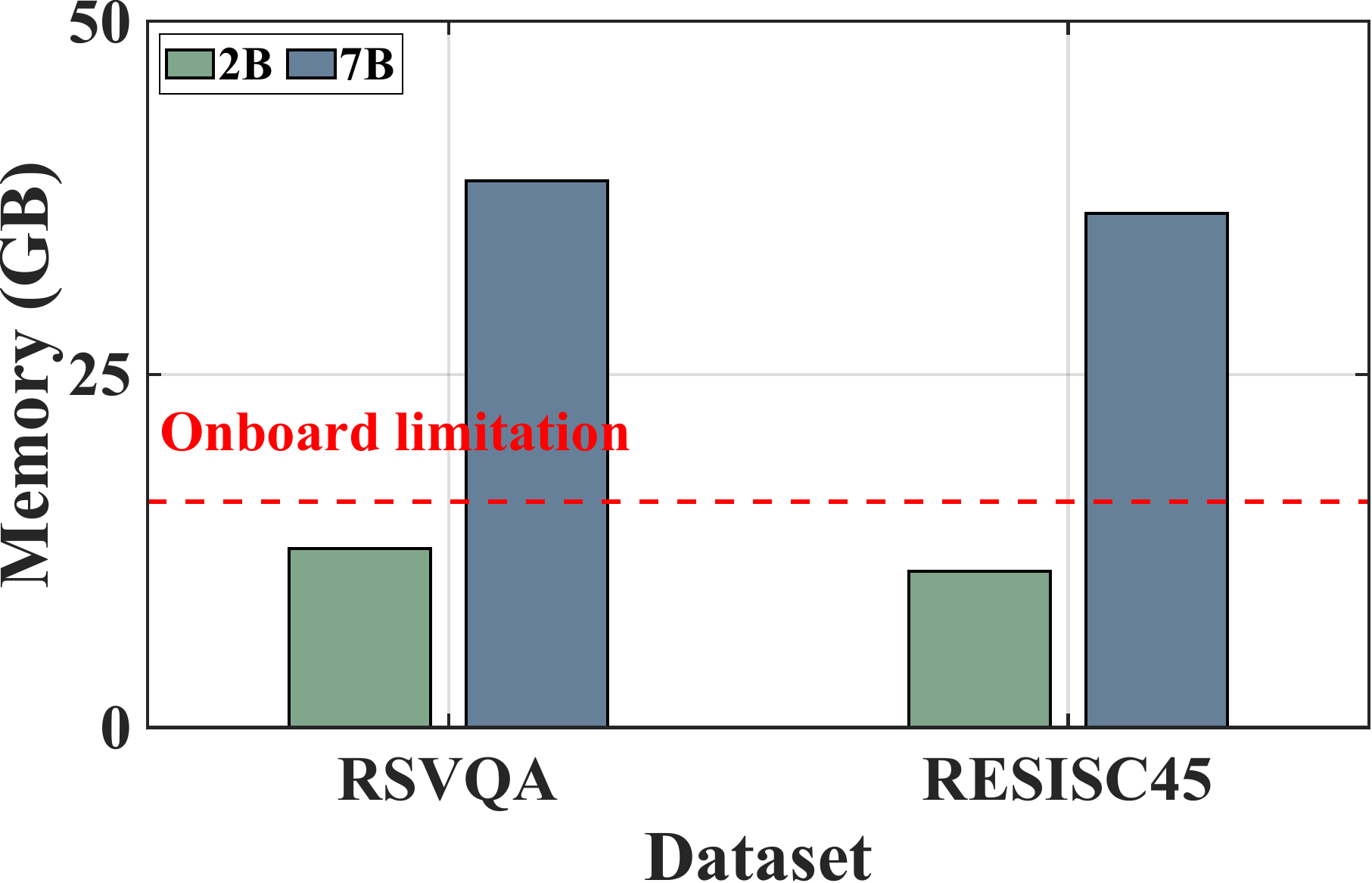}
        \caption{Peak memory usage.}
        \label{fig:moti-memo}
    \end{subfigure}
    \hfill
    \begin{subfigure}[b]{0.245\linewidth}
        \centering
        \includegraphics[width=\columnwidth, valign=c]{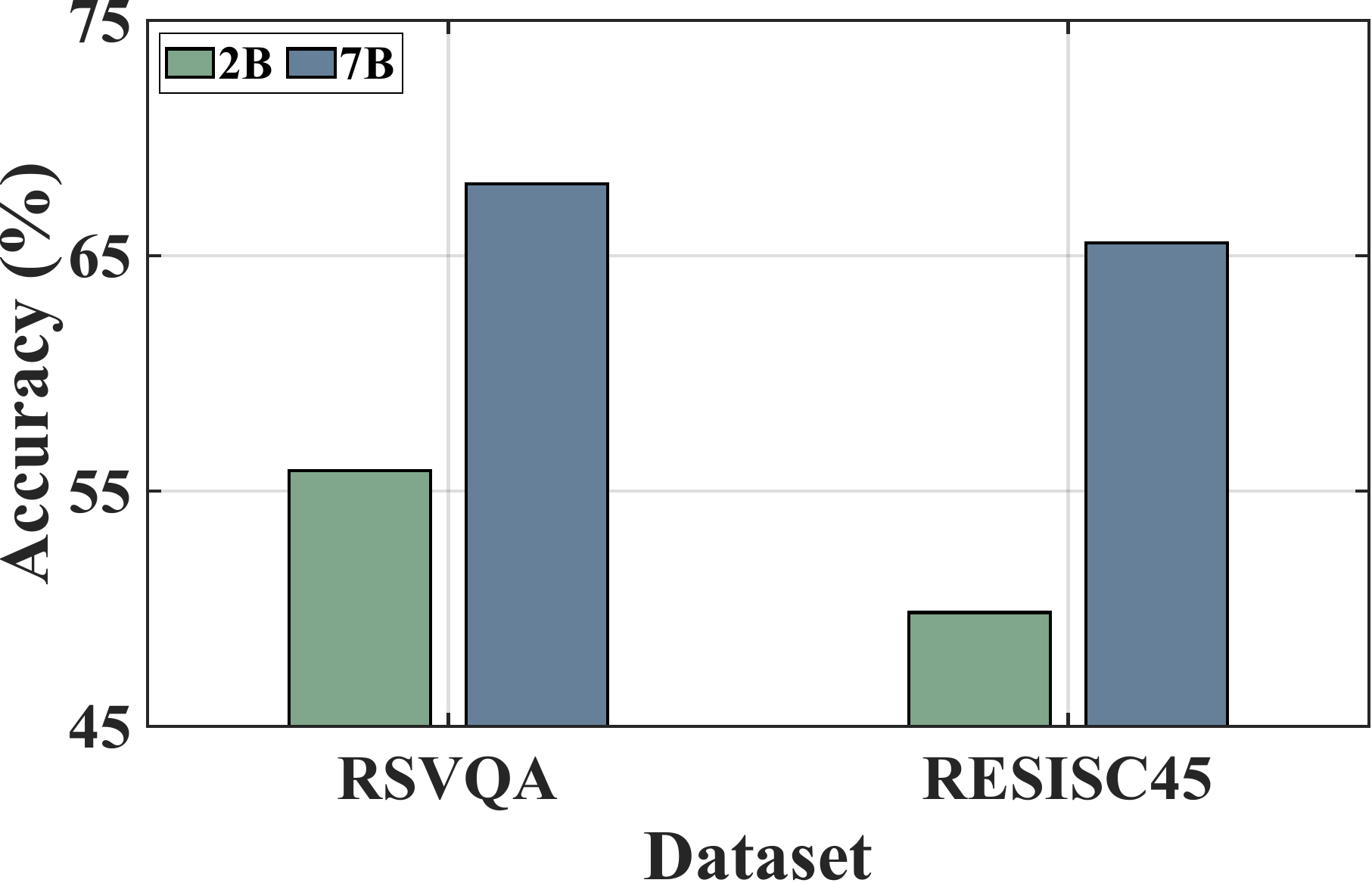}
        \caption{Accuracy.}
        \label{fig:moti-size}
    \end{subfigure}
    \hfill
    \begin{subfigure}[b]{0.2405\linewidth}
        \centering
        \includegraphics[width=\columnwidth, valign=c]{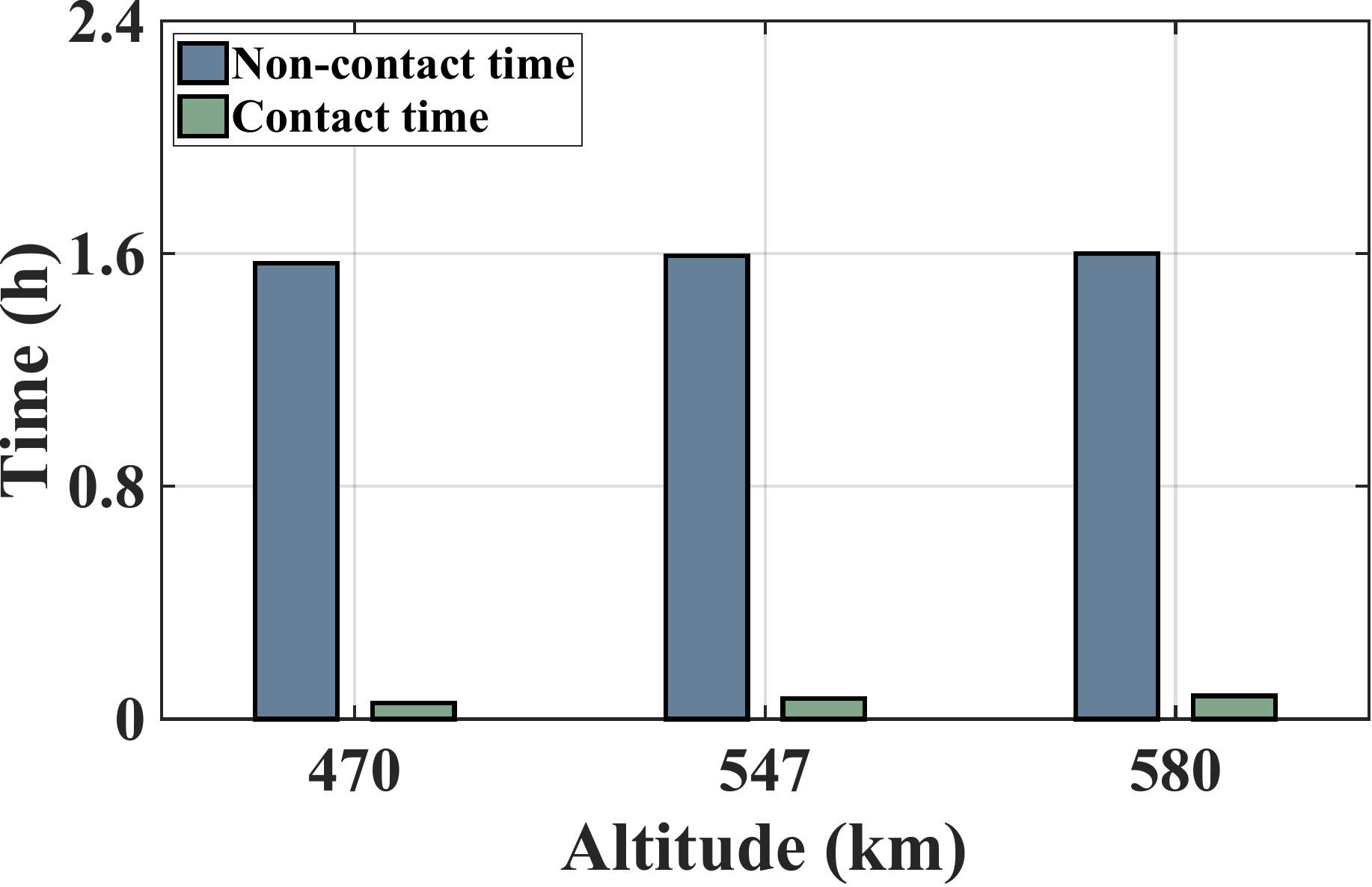}
        \caption{Contact time.}
        \label{fig:moti-inter}
    \end{subfigure}
    \hfill
    \begin{subfigure}[b]{0.25\linewidth}
        \centering
        \includegraphics[width=\columnwidth, valign=c]{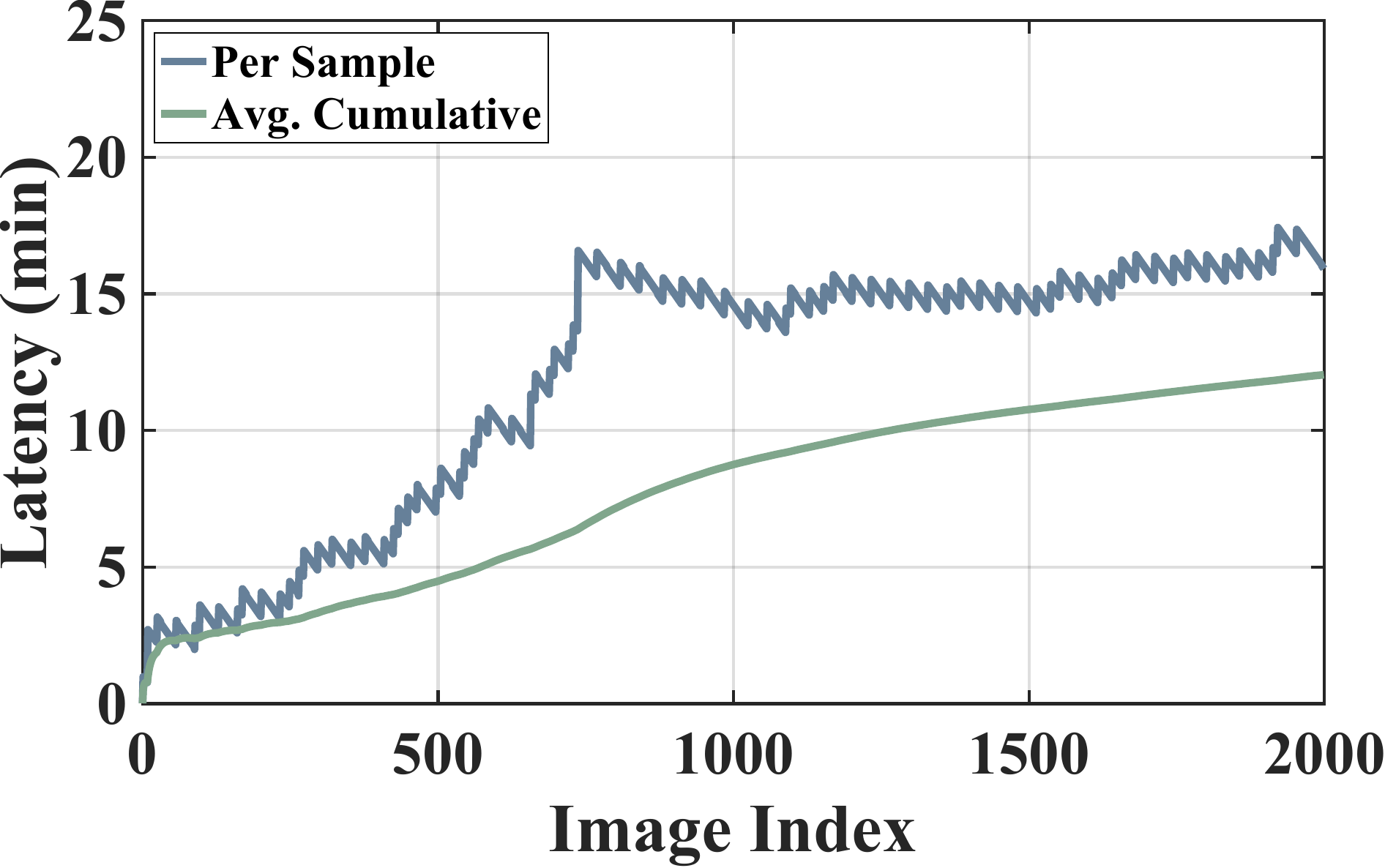}
        \caption{Transmission.}
        \label{fig:moti-lat}
    \end{subfigure}
    \caption{The the limited onboard resources (a) restrict the onboard generalization (b). The intermittent connection (c) and the limited transmission rate become a main bottleneck of task latency (d).}
    \label{fig:main}
\end{figure*}

\subsection{\yuxinzhangrev{Limited} Onboard Resources}

The computational resources on LEO satellites are severely limited compared to ground stations. While the latest LEO satellites have begun integrating \lizhrev{edge computing platforms (like NVIDIA Jetson)} to support onboard artificial intelligence \cite{satellitejetson}, the computational capacity of these platforms remains \lizhrev{moderate}. For example, the computing power of the \lizhrev{NVIDIA Jetson Orin NX is about 50 TFLOPS \cite{jetsontflops}.}
In contrast, GPUs available at ground stations, such as the NVIDIA A100 GPU, can achieve computing power up to 312 TFLOPS \cite{a100tflops}. This limitation renders LEO satellites incapable of running large-scale LVLMs. Furthermore, \lizhrev{onboard payloads and energy need to be prioritized for maintaining control systems, communicating with ground stations, remote image acquisition, and other critical satellite functions \cite{shergill2024energy, yuan2024heat, feng2021rflens}}, leaving minimal room for other applications, which further constrains the scale of onboard LVLMs. According to the scaling law \cite{kaplan2020scaling}, the generalization capability of LVLMs directly correlates with their size. Consequently, the inference accuracy of onboard LVLMs will inevitably underperform that of their larger-scale counterparts deployed on ground stations.

To validate this motivation, we conduct experiments using the RSVQA \cite{lobry2020rsvqa} and \lizhrev{RESISC45 \cite{cheng2017remote}} with the Qwen2-VL \cite{Qwen-VL}. We simulate onboard and ground-based LVLMs using 2B and 7B parameter models, respectively. \lizhrev{A Jetson Orin NX with 16GB memory is used to emulate the satellite environment.} \lizhrev{Figure \ref{fig:moti-memo} depicts the peak memory usage required by 2B and 7B LVLMs. The satellite environment only supports a 2B LVLM. Larger models can only be run at the ground station.} As shown in Figure \ref{fig:moti-size}, \lizhrev{for RSVQA,} the 2B model achieve an accuracy of 55.9\%, while the 7B model reach 68.0\%, a 12.1\% gap. \lizhrev{RESISC45 shows a similar result.} This finding confirms that relying solely on onboard models for local processing fails to meet user requirements for inference accuracy.

\subsection{\yuxinzhangrev{Brief} Satellite-GS \yuxinzhangrev{Contact}}

\begin{figure*}[t]
    \centering
    \includegraphics[width=1\linewidth]{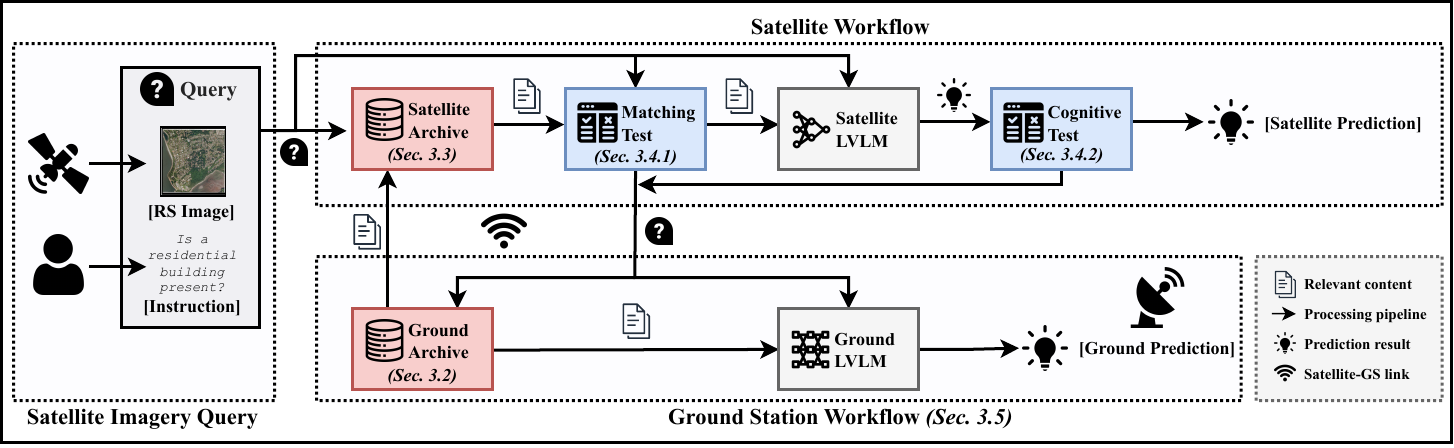}
    \caption{\lizhrev{Grace Overview. On the above, the satellite is equipped with a compact LVLM and a lightweight archive to process queries using onboard resources. On the below, the ground station features a comprehensive archive and a powerful LVLM, processing buffered queries and updating the satellite archive. ``Sec.'' indicates ``Section''.}}
    \label{fig:arch}
\end{figure*}

The satellite's data downlink capability cannot keep pace with its data acquisition capability. Due to their low orbital altitude and high-velocity relative to Earth’s rotation, LEO satellites experience intermittent connectivity with ground stations, where the duration of network availability is significantly shorter than outages. Based on Starlink’s LEO constellation configuration \cite{starlinkorb}, we calculate the average contact time between LEO satellites and ground stations at different orbital altitudes. As illustrated in Figure \ref{fig:moti-inter}, an LEO satellite completes an Earth orbit in approximately 95 minutes, with only about 5 minutes of viable communication time per pass. This intermittency poses challenges for directly applying existing collaborative inference frameworks designed for large language models, as these frameworks assume persistent network connectivity among computing nodes --- an assumption easily satisfied in terrestrial scenarios but invalid for LEO environments. Thus, novel methods are required to adapt to intermittent connectivity in satellite-ground systems. \lizhrev{Therefore, LEO satellites must rely on their own capabilities to complete inference tasks for the vast majority of the time.}

LEO satellites are equipped with high-speed sensors capable of collecting vast amounts of remote sensing data. Remote sensing data can be acquired at any time during the satellite's flight. However, their data transmission bandwidth to ground stations is disproportionately limited. 
\lizhrev{For instance, WorldView-3 can acquire up to 680,000 square kilometers per day; each 13 kilometers $\times$ 112 kilometers scene is roughly 40 Gb after compression. Its fastest X-band channel tops out at 1200 Mbps \cite{worldview}. If the sensor were operated at maximum duty cycle, the satellite-GS link cannot complete the transmission of all data within the brief connectivity window.}
When multiple tenants share the satellite’s bandwidth, the allocation per user becomes even more constrained. Ground-centric inference approaches face severe bottlenecks in such scenarios due to insufficient data throughput \cite{singh2021community, pan2023pmsat}.

We simulate a scenario using the high-resolution RSVQA dataset to demonstrate the impact of bandwidth limitations. Assuming a LEO satellite captures one image every 2 seconds and transmits data compressed by \textit{zlib} to the ground at 30 Mbps during 3-second intervals every minute, we measured the latency between image acquisition and ground processing over time. As shown in Figure \ref{fig:moti-lat}, images captured later experience significantly longer transmission delays, with both average and maximum latency exhibiting an upward trend. This occurs because the transmission rate cannot keep pace with data collection, leading to an accumulation of onboard data and progressively worsening processing delays. Such bottlenecks critically degrade the efficiency of remote sensing inference tasks.

\section{System Design}
\label{sec:system-design}

\subsection{Overview}

\lizhrev{In this section, we present Grace, a satellite-ground collaborative LVLM inference system, as illustrated in Figure \ref{fig:arch}. In our setting, each query \(Q\) comprises a remote sensing image \(Q_M\) and a natural language instruction \(Q_I\), i.e., \(Q = \{Q_M, Q_I\}\). Grace aims to interpret the image \(Q_M\) according to the instruction \(Q_I\), delivering accurate and efficient inference within the satellite’s resource constraints. To address the challenges introduced in Section \ref{sec:challenges-and-motivation}, We will provide a detailed introduction to Grace through the following sections:}

\begin{itemize}
    \item The \textit{ground archive} (Section \ref{sec:dynamic-knowledge-archive}) is deployed to equip the LVLM with domain-specific knowledge and enhance its inference capabilities. This section introduces the data structure and the retrieval scheme of the archive. 
    \item \lizhrev{The \textit{satellite archive} (Section \ref{sec:hierarchical-transmission-mechanism}) is integrated with dynamic adaptation algorithm, which allows the satellite archive to automatically update its content in response to changes in mission requirements and RS image content, ensuring adaptability to new scenarios.}
    \item To reduce data transmission requirements, a \textit{task dispatcher} (Section \ref{sec:task-dispatcher}) strategically assigns inference tasks. Most tasks are processed onboard, significantly reducing the number of samples that must be transmitted to the ground station.
    \item Leveraging the disparity in computational resources, a high-performance LVLM with strong generalization capabilities is deployed at the \textit{ground station} (Section \ref{sec:ground-inference}), providing enhanced support for complex tasks.
\end{itemize}


\subsection{Ground Archive}
\label{sec:dynamic-knowledge-archive}

A specially constructed external knowledge archive is critical to flexibly update the external knowledge for LVLMs and enhance inference accuracy. In our system, the LEO satellite and the ground station maintain an archive as an external knowledge repository for the LVLM. 
\lizhrev{We will introduce the fetching method and the search result for our archive in this Section. Then we will describe the corresponding adoptive update algorithm at Section \ref{sec:hierarchical-transmission-mechanism}.}

\subsubsection{Fetching Method}

The archive stores a large collection of remote sensing images \(\{R_{M_1}, R_{M_2}, R_{M_3}, \ldots\}\). Formally, for the $j$-th remote sensing image \(R_{M_j}\), there is a set of instruction-answer pairs \(\{\langle R_{I_{j,1}}, R_{A_{j,1}}\rangle, \langle R_{I_{j,2}, R_{A_{j,2}}\rangle}, \langle R_{I_{j,3}}, R_{A_{j,3}}\rangle, \ldots\}\). \lizhrev{Each ground-truth has been carefully annotated by human experts. The data structure can be formulated as Equation \ref{equ:archive}.
\begin{equation}
\left\{ 
\begin{array}{l}
R_{M_1}: \left\{ \langle R_{I_{1,1}}, R_{A_{1,1}} \rangle, \langle R_{I_{1,2}}, R_{A_{1,2}} \rangle, \ldots \right\} \\
R_{M_2}: \left\{ \langle R_{I_{2,1}}, R_{A_{2,1}} \rangle, \langle R_{I_{2,2}}, R_{A_{2,2}} \rangle, \ldots \right\} \\
R_{M_3}: \left\{ \langle R_{I_{3,1}}, R_{A_{3,1}} \rangle, \langle R_{I_{3,2}}, R_{A_{3,2}} \rangle, \ldots \right\} \\
\ldots
\end{array} 
\right\}.
\label{equ:archive}
\end{equation}}
The objective of the archive is to retrieve the content most relevant to a query \(Q\). Notably, each query \(Q\) is multi-modal, comprising a newly captured satellite image \(Q_M\) \lizhrev{and a per-defined instruction $Q_I$}. Therefore, we employ separate modules for vision and text queries that jointly identify relevant content. 

\paragraph{Vision Query Module}

The vision query module uses a vision embedding model \(E_v\) to convert images into feature vectors. Specifically, every remote sensing image \(R_{M_j}\) in the knowledge base is embedded and normalized to a unit vector: \(e^R_{M_j} = {E_v(R_{M_j})}/{\| E_v(R_{M_j}) \|}.\) These normalized vectors are stacked to form a matrix:
\begin{equation}
   A_M = 
   \begin{bmatrix}
       e^R_{M_1},e^R_{M_2},e^R_{M_3},\ldots
   \end{bmatrix},
   \label{equ:am}
\end{equation}
which is stored in the archive for subsequent retrieval.

When a query image \(Q_M\) arrives, the embedding model \(E_v\) produces a normalized query vector \(e^Q_{M} = {E_v(Q_{M})}/{\| E_v(Q_{M}) \|}.\) We then compute the cosine similarity of \(e^Q_{M}\) with each column in \(A_M\): \lizhrev{
\begin{equation}
    \mathrm{Sim}_M = {A_M}^T \cdot e^Q_M,
\end{equation}
yielding similarity scores between \(M_Q\) and stored images.}

\paragraph{Instruction Query Module}

Natural-language instructions in the knowledge base are processed in a similar fashion using a text embedding model \(E_t\). However, we first de-duplicate the instructions to minimize computation and storage, as different images might share the same instruction. After removing duplicates, we obtain a unique instruction set \(\{R_{I_1}, R_{I_2}, R_{I_3}, \ldots\}\) and maintain a mapping function \(F\) that memories which instructions belong to which images. \lizhrev{``$j \in F(i)$'' indicates that the instruction set corresponding to the j-th image $R_{M_j}$ contains the instruction $R_{I_i}$.}

\lizhrev{Next, each unique instruction \(R_{I_j}\) is embedded and normalized: \(e^R_{I_j} = {E_t(R_{I_j})}/{\|E_t(R_{I_j})\|}.\) These vectors form a matrix \(A_I\) like \(A_M\) in Equation \ref{equ:am}. For the query instruction \(Q_I\), \(E_t\) generates a normalized embedding \(e^Q_I = {E_t(Q_I)}/{\|E_t(Q_I)\|}.\) We then compute the similarity vector:
\begin{equation}
\mathrm{Sim}_I =  {A_I}^T \cdot e^Q_I,
\end{equation}
where each element indicates how well \(I_Q\) matches each instruction in the knowledge base.}

\paragraph{Data Fusion Module}

Although one image may have multiple instructions, including all of them in the prompt can overwhelm the LVLM and degrade inference performance. Excessive, irrelevant instructions can increase computation and dilute the essential context needed by the LVLM \cite{asai2023self}. To balance coverage and relevance, the data fusion module aggregates \(\mathrm{Sim}_M\) and \(\mathrm{Sim}_I\) to produce the integrated retrieval results. For every remote sensing image $R_{M_i}$ in the archive, the data fusion module consults the mapping function \(F\) to find the instruction with the highest similarity to \(Q_I\):
\begin{equation}
\mathrm{Sim}_i = \mathrm{Sim}_{M_i} + \max_{j \in F(i)} \mathrm{Sim}_{I_j}.
\end{equation}
The data fusion module picks the single relevant instruction from the archive that best aligns with \(Q_I\). Finally, it returns the top-\(K\) relevant records.

\begin{figure}[t]
    \centering
    \includegraphics[width=\columnwidth]{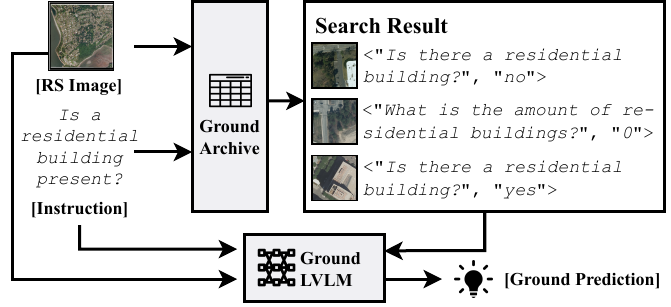}
    \caption{\lizhrev{Ground-station inference process. Relevant data for the query is retrieved by the ground archive. The ground LVLM takes both the query and the search result as input to generate the inference outcome.}}
    \label{fig:infer}
\end{figure}

\subsubsection{Search Result Structure}

\begin{figure*}[t]
    \centering
    \includegraphics[width=\linewidth]{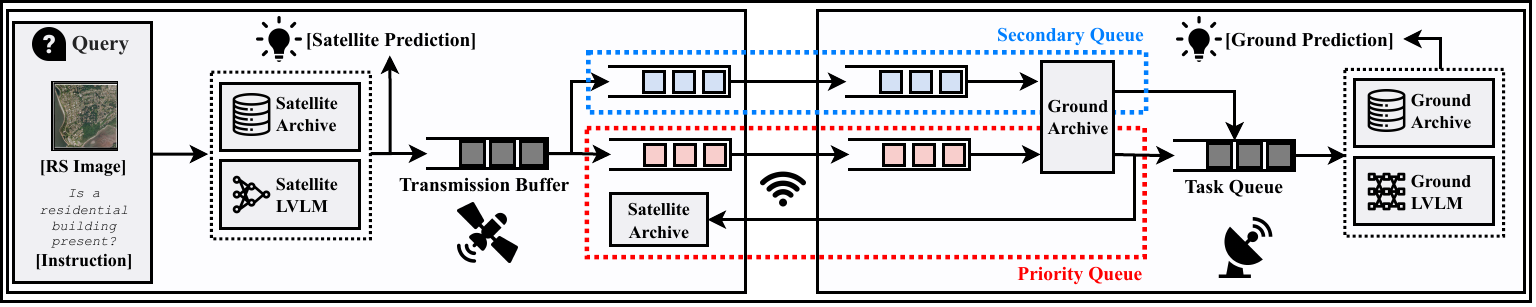}
    \caption{\lizhrev{The hierarchical transmission mechanism. The LEO satellite side (left) and the ground station side (right). The priority queue will be transmitted to the ground station first, and its retrieval results will be sent back for updating the satellite archive. The secondary queue will be transmitted after the completion of the priority queue transmission.}}
    \label{fig:flow}
\end{figure*}

The retrieved relevant content \(C\) is a set of records \(\{C_1, C_2, \ldots\}\). The $i$-th record \(C_i\) contains following items:

\begin{itemize}
    \item \lizhrev{A relevant remote sensing image \(C_{i_M}\)} from the archive that is similar to the query image \(Q_M\) \lizhrev{and a corresponding similarity score \(S_{i_M}\) measuring the imaginary relevance.}
    \item \lizhrev{A relevant instruction \(C_{i_I}\)} closely related to the query instruction \(Q_I\) \lizhrev{and a corresponding  similarity score \(S_{i_I}\) measuring the textual relevance.}
    \item \lizhrev{A human-annotated ground-truth answer \(C_{i_G}\)} for \(\{C_{i_M}, C_{i_I}\}\). \lizhrev{This ground-truth helps guide the LVLM by showing the appropriate expected answer for tasks similar to the query.}
\end{itemize}

\lizhrev{Figure \ref{fig:infer} presents an example of the search process. Each row in the ``Search Result'' retrieved from the archive represents a record \(C_i\), which contains an image  \(R_{i_M}\), an instruction \(R_{i_I}\), and a ground-truth \(R_{i_G}\). The search result \(C\), along with the query, will be fed into the LVLM for inference to generate the final answer.}

\subsection{Satellite Archive}
\label{sec:hierarchical-transmission-mechanism}

The LEO satellite archive must be highly optimized for the satellite’s resource-constrained environment. In practice, the LEO satellite archive is a subset of the ground archive that includes only the data most relevant to the satellite’s current tasks. 

\lizhrev{The satellite archive can provide sufficient relevant information to the satellite LVLM, even with limited content size, for two primary reasons: 1) Predictable orbital path: The operational orbit of satellites is fixed and predictable, which limits the geographical locations of captured images. It is possible to know in advance what locations can be imaged at certain time. This predictability ensures that the archive contains a structured and manageable set of data related to specific geographic areas. 2) Temporal stability of land: Most terrestrial features are stable over time. Urban infrastructures and agricultural landscapes typically undergo minimal changes over periods of multiple days. This slow-changing characteristic of land dynamics means that images of the same location often maintain consistent content over time. As a result, even a limited archive can offer valuable and relevant knowledge for inference purposes without needing extensive updates or large volumes of new imagery.}
Consequently, a carefully chosen subset of the ground archive can support stable onboard inference while minimizing storage overhead.

However, determining which portion of the ground archive to upload to the satellite is non-trivial. Moreover, if the satellite’s mission scope changes over time, the archive must also adapt. To address these challenges, we design a dynamic adaptation algorithm. \lizhrev{This algorithm consists of two components: a replace module \lizhrev{running in the satellite archive} and a hierarchical transmission mechanism.}

\subsubsection{Replace Module}

The replace module tracks the usage patterns of each remote-sensing image in the satellite archive. Every time a query is processed onboard, the top-\(K\) retrieved images are moved to the front of a queue, indicating that they have been recently accessed. Conversely, images that seldom appear in the results move toward the back of this queue over time.

When the archive requires an update, the replace module removes entries from the tail of the queue (those least recently used). These outdated or unused items are discarded to free space. Newly introduced images and instructions are then loaded into the archive and placed at the front of the queue, ensuring that high-demand data are always prioritized.

\subsubsection{Hierarchical Transmission Mechanism}



\lizhrev{LEO satellites move at high speed relative to the ground, resulting in intermittent and brief communication windows between the satellite and the ground station. On average, each window lasts only about five minutes, followed by roughly ninety minutes of no connectivity. During these brief windows, the satellite can transmit tasks accumulated in its cache buffer to the ground station. Simultaneously, the ground station processes these queries using its comprehensive archive and sends back relevant content to update the satellite archive. However, the volume of data transmitted from the ground to the satellite is constrained by the limited bandwidth, which is typically smaller than the reverse direction \cite{lin2025leo}. This asymmetry necessitates a mechanism to prioritize critical data transfers during the short communication windows.

To address this challenge, we introduce a hierarchical transmission mechanism based on the operational characteristics of LEO satellites. The satellite operates over prolonged durations, during which its transmission cache may accumulate a large number of queries pending ground-based inference. Notably, newly appended queries in the buffer correspond to recent tasks that the satellite could not resolve locally. Prioritizing these recent queries for archive updates can reduce the volume of data transmitted from the ground to the satellite. The ground station only needs to transmit the relevant content for the priority queue, rather than for all accumulated queries. The volume of data transmitted from the satellite to the ground remains unaffected, as all queries must eventually be processed for inference. Based on this rationale, we split the transmission buffer into two queues: 1) a priority queue containing the most recent $N_{\mathrm{mp}}$ queries, and 2) a secondary queue holding all other queries. This design optimizes the use of the asymmetric communication channel, ensuring efficient utilization of the limited ground-to-satellite bandwidth while maintaining the integrity of the inference pipeline.}

\paragraph{Priority Queue}

The priority queue is given preferential access to communication resources on the satellite and ground station. Specifically, 1) on the satellite: The priority queue must finish sending its queries before any queries from the secondary queue can be transmitted. 2) On the ground station: The priority queue’s queries must be processed first before the ground station handles any secondary queue requests.

When a communication window opens, the satellite transmits all queries in the priority queue to the ground station. Suppose the priority queue contains \(N_{\mathrm{prior}}\) queries (\(N_{\mathrm{prior}} \leq N_{\mathrm{mp}}\)). The ground station uses its ground archive to retrieve relevant content for each query, producing a total of \(K \times N_{\mathrm{prior}}\) relevant records. After de-duplication, the ground station sends metadata (including remote sensing image IDs) for these records to the satellite. The satellite compares incoming IDs against its onboard archive. It identifies any missing data and sends a request back to the ground station for those missing content. The ground station then transmits all missing images, corresponding instructions, and ground-truths to the satellite, which inserts the newly received data into its satellite archive. This process ensures that the satellite archive is updated with the most relevant information for current tasks.

\paragraph{Secondary Queue}

Queries in the secondary queue are generally older and are therefore addressed only after the priority queue has been fully processed. Secondary queue transmission begins when the satellite confirms that all priority queue transmissions are complete. The secondary queue is transmitted in chunks because satellite communication links can fail unexpectedly. In contrast to the priority queue, the ground station does not immediately process these secondary queries. Instead, it defers them until no further priority tasks remain. At this point, it retrieves the relevant records from the ground archive only for local inference. Unlike priority queue updates, the ground station does not return relevant content to the satellite. Once the ground station receives the previous chunk of data, the satellite removes the delivered queries from the local secondary queue and proceeds to transmit the next batch of queries. Figure \ref{fig:flow} shows the task flow between the LEO satellite and the ground station.

\begin{algorithm}[t]
\caption{Task Dispatcher}
\label{alg:dispatch}
\begin{algorithmic}[1]
\Require Query $Q = \{Q_M, Q_I\}$, relevant content $C$, satellite LVLM $M_{sat}$, thresholds $T_M$, $T_I$, $T_K$, $T_{\text{Conf}}$.
\Ensure Decision: Accept the onboard inference result or transmit to ground station.
\Function{Dispatch}{$Q$}    
    \State \textbf{Stage 1: Matching Test}
    \State $C' \gets \emptyset$
    \For{each record $R_i \in C$}
        \If{$S_{i_M} \geq T_M$ \textbf{and} $S_{i_I} \geq T_I$}
            \State $C' \gets C' \cup \{R_i\}$
        \EndIf
    \EndFor
    
    \If{$|C'| < T_K$}  \Comment{Insufficient relevant records}
        \State \text{Cache $Q$ for transmission}
        \State \Return $\text{Transmit}$
    \EndIf
    \State $\text{Pred}, \text{Prob} \gets M_{sat}(Q, C')$

    \State \textbf{Stage 2: Cognitive Test}
    \State $\text{Conf} \gets 0$
    \For{$i \gets N_{\text{inp}}$ \textbf{to} $N_{\text{inp}} + N_{\text{gen}}$}
        \State $\text{Conf} \gets \text{Conf} + \ln(\text{Prob}_i)$
    \EndFor
    \State $\text{Conf} \gets \exp\left({\text{Conf}}/{N_{\text{gen}}}\right)$  
    \If{$\text{Conf} < T_{\text{Conf}}$}  \Comment{Insufficient confidence}
        \State \text{Cache $Q$ for transmission}
        \State \Return $\text{Transmit}$
    \Else
        \State \Return $\text{Accept}$  \Comment{Accept onboard inference result}
    \EndIf
\EndFunction
\end{algorithmic}
\end{algorithm}

\subsection{Task Dispatcher}
\label{sec:task-dispatcher}

\lizhrev{When an LEO satellite captures a remote sensing image \(Q_M\), the onboard inference module is immediately activated to process it according to a per-defined instruction \(Q_I\). The satellite archive queries to retrieve relevant content $C$ that corresponds to \(Q\). However, }
due to limited computing and storage resources, the satellite archive and the LVLM are constrained in size. Consequently, the onboard inference may not always provide correct answers. Therefore, we propose a task dispatch algorithm to decide which tasks should be kept locally for inference and which tasks cannot be completed on the satellite and must be sent to the ground station for processing.

\lizhrev{The core requirement of the dispatch algorithm is to assess the confidence level of the onboard inference system in accurately completing the inference task. If the confidence level is sufficiently high, the inference can be completed on the satellite; otherwise, the raw image needs to be transmitted to the ground station for further processing. The task dispatch consists of two stages. The first stage is a matching test conducted before inference, and the second stage is a cognitive test carried out after inference. If either test is not passed, the query must be sent to the ground station for processing. Otherwise, the inference can be completed on the satellite. Algorithm \ref{alg:dispatch} shows the entire dispatching procedure.}

\subsubsection{Matching Test}

Before inference begins, the dispatcher conducts a matching test to evaluate whether the retrieved relevant content $C$ adequately supports the inference. The matching test proceeds in two steps. 

1) The dispatcher filters out low-relevance records from the retrieved relevant content. Each relevant record \(R_i\) is associated with two similarity scores: the image similarity $S_{i_M}$ and the instruction similarity $S_{i_I}$. We define two thresholds, \(T_M\) for image similarity and \(T_I\) for instruction similarity. Any record whose image or instruction similarity falls below the corresponding threshold is discarded from the relevant content. 

2) The dispatcher checks whether the remaining relevant content \(C'\) has at least \(T_K\) records (where \(T_K \leq K\)). This lower bound ensures that the LVLM has enough context to complete a reliable inference. If \(\lvert C' \rvert < T_K\), we conclude that the satellite archive lacks the necessary information for this query. In that case, the query \(Q\) is cached for transmission to the ground station, where a more comprehensive archive and a more capable LVLM can handle the inference task.

If the filtered relevant content \(C'\) passes the matching test, it proceeds to the satellite inference pipeline. The content in \(C'\) is integrated with the task \(Q\) during the prompt assembly. \lizhrev{Lines 2 to 13 of Algorithm \ref{alg:dispatch} illustrate the procedure of the matching test.}

\subsubsection{Cognitive Test}

Although relevant content helps compensate for the LVLM’s limited model size and generalization capabilities onboard the satellite, the system still cannot fully trust the final result without an additional verification step. We introduce a cognitive test to assess the credibility of the LVLM’s answer.

During LVLM inference, we track each newly generated token \(x_i\) and record its probability \(P(x_i \mid x_{<i})\), where \(x_{<i} = \{x_0, x_1, \ldots, x_{i-1}\}\). A high probability implies that the LVLM is confident about the token choice. Since the final output consists of multiple tokens, we compute the geometric mean of the probabilities of all newly generated tokens as an overall confidence score:
\begin{equation}
   \mathrm{Conf} = \exp\Biggl(\frac{1}{N_{\mathrm{gen}}}\sum_{i=N_{\mathrm{inp}}}^{N_{\mathrm{inp}} + N_{\mathrm{gen}}} \ln P(x_i \mid x_{<i})\Biggr),
\end{equation}
where \(N_{\mathrm{gen}}\) is the length of the output response and \(N_{\mathrm{inp}}\) is the size of input tokens. We then compare this \(\mathrm{Conf}\) value against a threshold \(T_{\mathrm{Conf}}\). If \(\mathrm{Conf} < T_{\mathrm{Conf}}\), it indicates the satellite LVLM is not sufficiently confident in its inference. The system discards the result and places the query \(Q\) in a transmission cache to await communication to the ground station. Otherwise, the result is accepted as the final answer, and no further transmission is needed. \lizhrev{Lines 14 to 25 of Algorithm \ref{alg:dispatch} illustrate the procedure of the cognitive test.}

\subsection{Ground Inference}
\label{sec:ground-inference}

\begin{figure}[t]
    \centering
    \includegraphics[width=0.95\columnwidth]{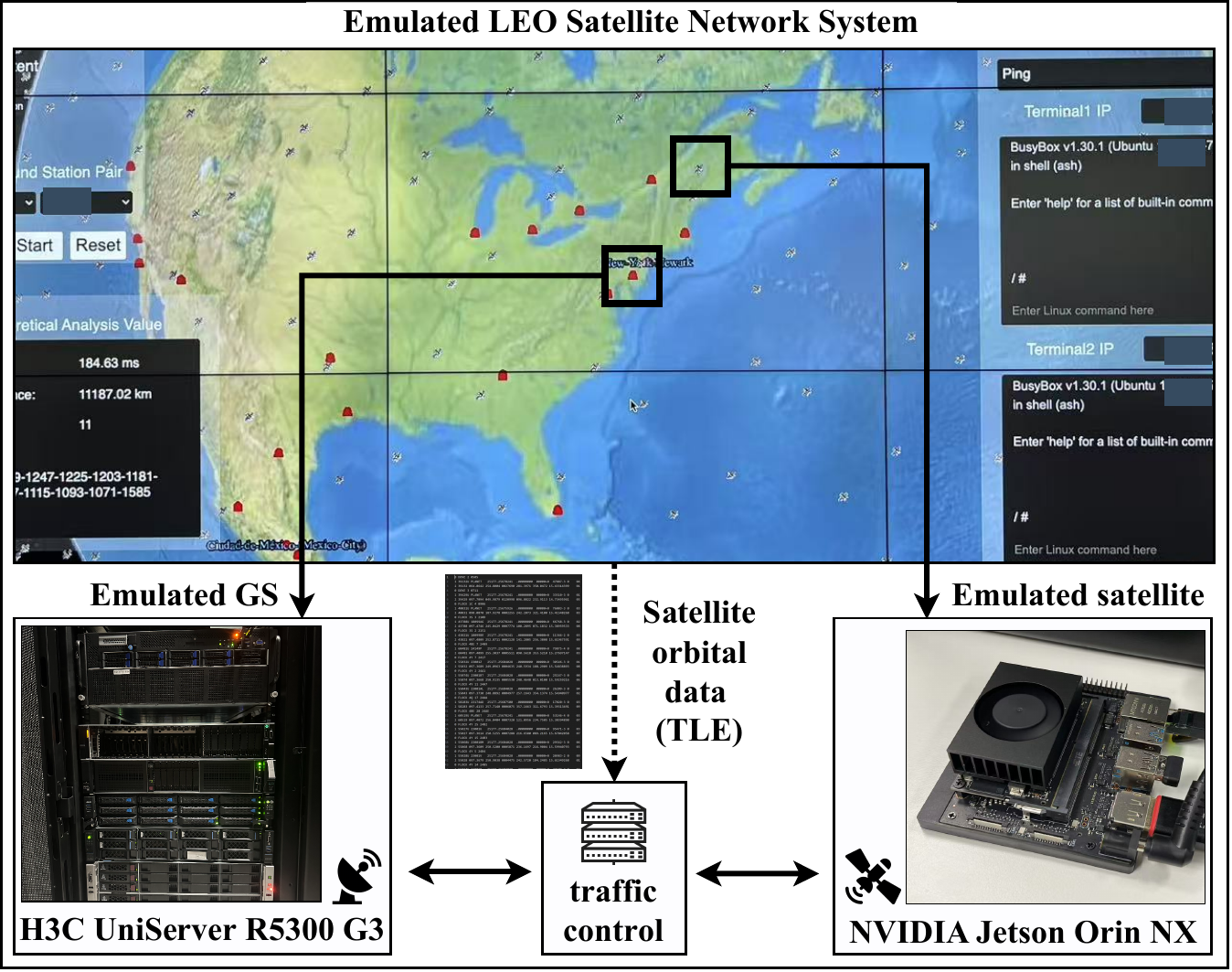}
    \caption{Grace testbed.}
    \label{fig:impl}
\end{figure}

\begin{figure*}[t]
    \centering
    \begin{subfigure}[b]{0.245\linewidth}
        \centering
        \includegraphics[width=\columnwidth]{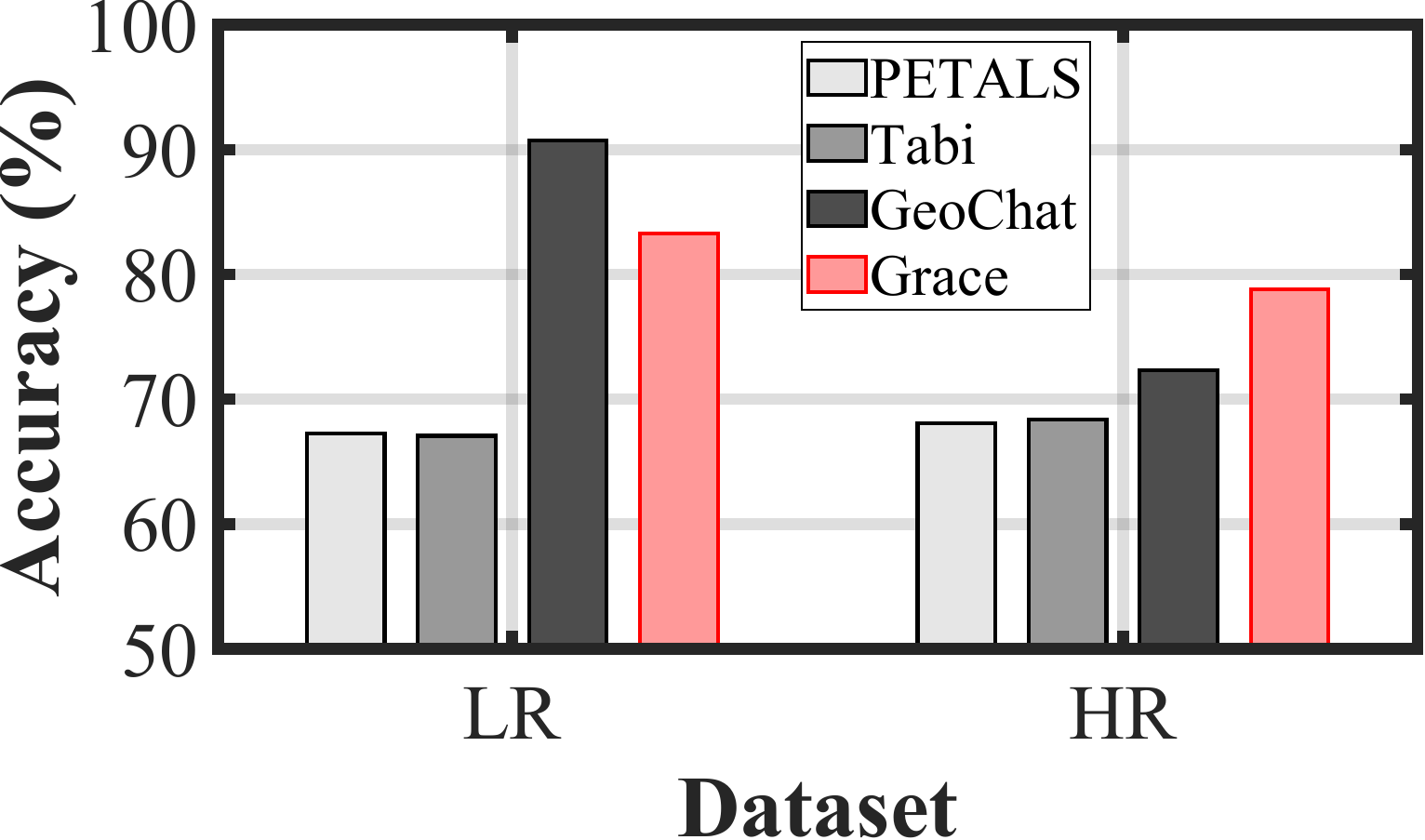}
        \caption{Test accuracy.}
        \label{fig:main-acc}
    \end{subfigure}
    \hfill
    \begin{subfigure}[b]{0.245\linewidth}
        \centering
        \includegraphics[width=\columnwidth]{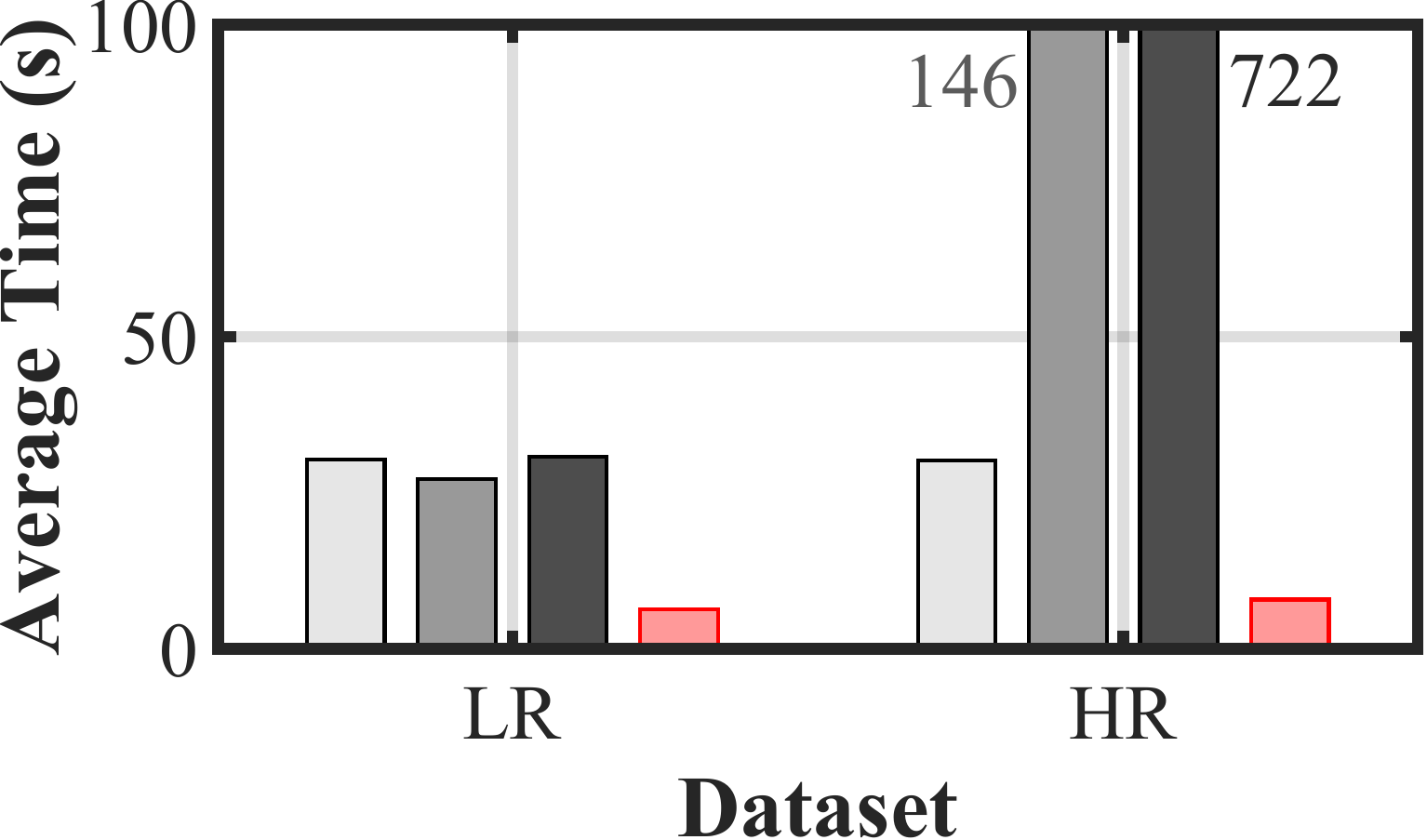}
        \caption{Average time.}
        \label{fig:main-avg}
    \end{subfigure}
    \hfill
    \begin{subfigure}[b]{0.245\linewidth}
        \centering
        \includegraphics[width=\columnwidth]{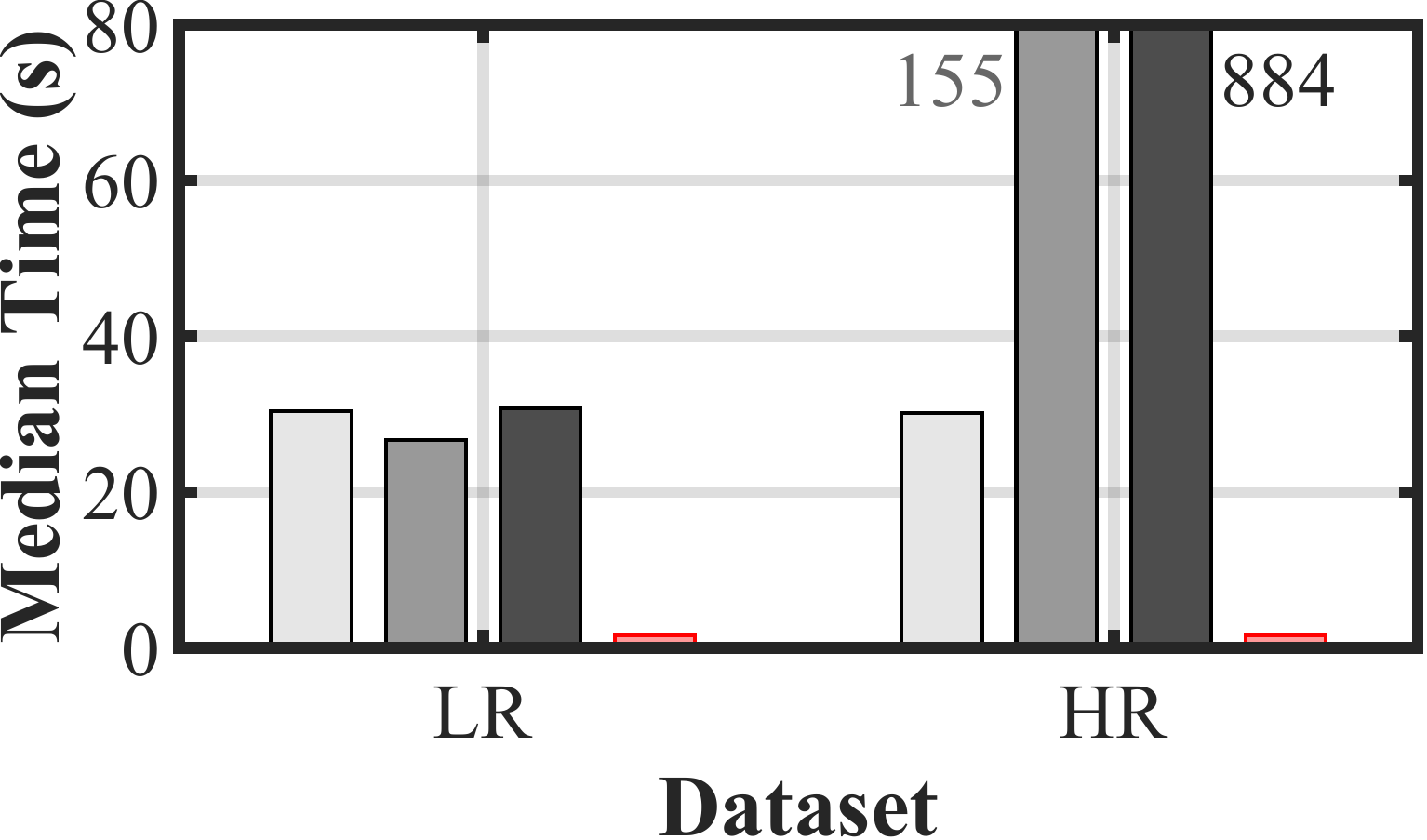}
        \caption{Median time.}
        \label{fig:main-med}
    \end{subfigure}
    \hfill
    \begin{subfigure}[b]{0.245\linewidth}
        \centering
        \includegraphics[width=\columnwidth]{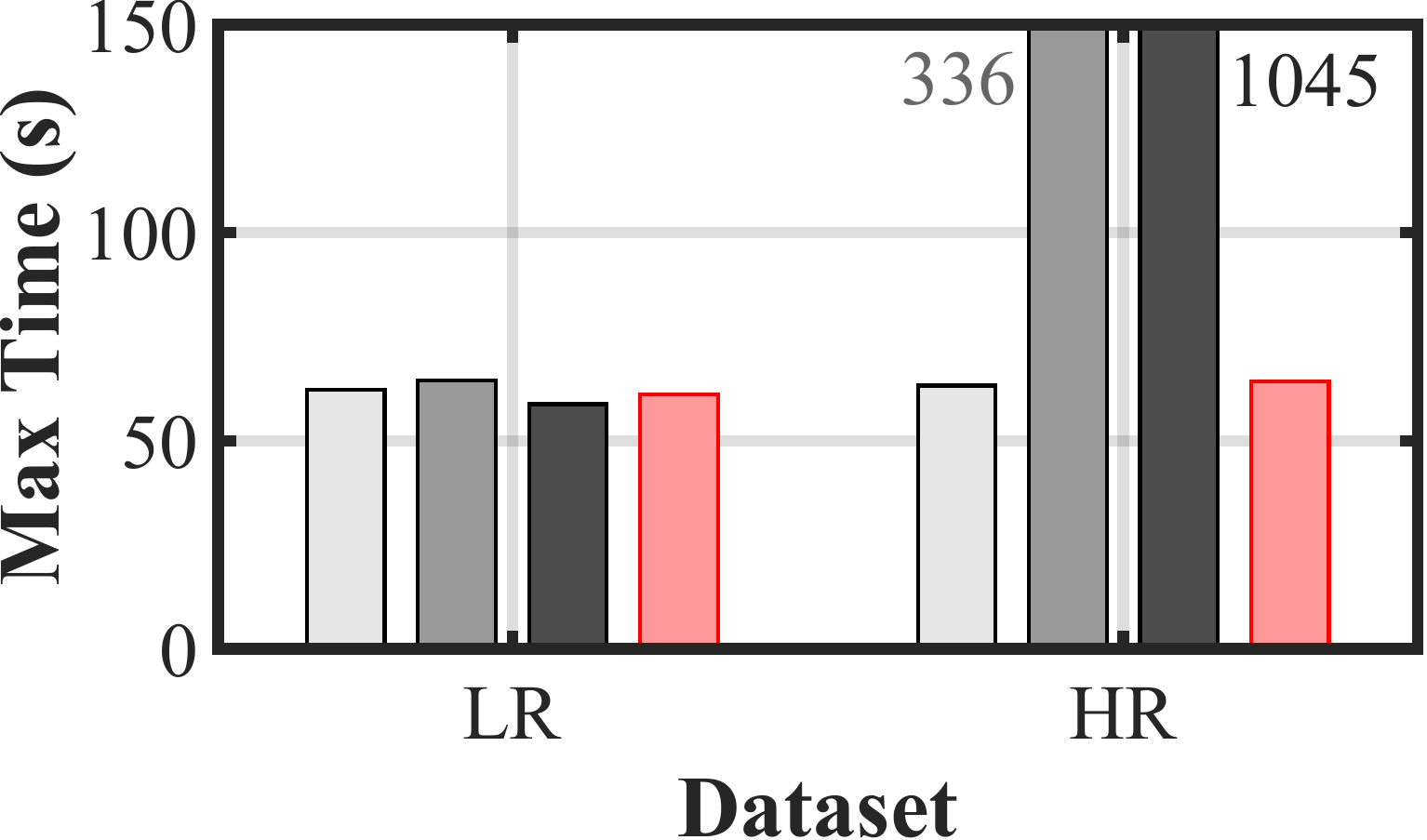}
        \caption{Max time.}
        \label{fig:main-max}
    \end{subfigure}
    \caption{Test accuracy and inference time on RSVQA LR and HR datasets. \lizhrev{The numbers beside bars in the upper part of the figure indicate actual heights of bars.}}
    \label{fig:main}
\end{figure*}

Unlike the LEO satellite inference pipeline, the ground station does not perform a matching test on the retrieved relevant content. For each query received --- whether from the priority queue or the secondary queue --- the ground archive returns \(K\) relevant records. All \(K\) records are assumed relevant and thus automatically included as input to the ground LVLM.

Each query and its corresponding relevant content are appended to a task queue. The ground station’s inference module activates whenever the task queue is non-empty. By separating the retrieval process from the LVLM inference, the ground station can rapidly respond to newly arrived priority queries during the short communication window, allowing more data to be downloaded from the satellite before the link is lost.

The LVLM inference at the ground station mirrors the satellite’s process. The cognitive test is omitted because the ground station typically has more robust and comprehensive resources. Therefore, the inference result becomes final without additional validation. 


\section{Implementation}
\label{sec:implementation}

In this section, we introduce the implementation of Grace, then we give the experiment setup.

\subsection{Implementing Grace}

The prototype of Grace is illustrated in Figure \ref{fig:impl}. The \lizhrev{ground station} is deployed in an H3C UniServer R5300 G3 equipped with eight NVIDIA GeForce RTX 3090 GPUs, dual Intel Xeon Silver 4210R processors (10 cores, 2.84 GHz each), and 8 $\times$ 32 GB DDR4 RAM, running Ubuntu 18.04.6 LTS. The software stack includes Python 3.12.4 and PyTorch 2.5.1.
\lizhrev{The satellite is deployed in an NVIDIA Jetson Orin NX equipped with an Arm Cortex-A78AE v8.2 CPU (8 cores, 2 GHz each), and 16 GB LPDDR5 RAM, running Ubuntu 22.04.5 LTS. The software stack includes Python 3.10.16 and PyTorch 2.7.0.}
\lizhrev{The testbed incorporates real-world orbital data (Two Line Element) of Planet’s Dove satellite constellation to facilitate real-time trajectory computation and network routing. The satellite-ground network condition is emulated through \textit{tc} \cite{beshay2015fidelity}, thereby establishing dynamic connectivity between ground stations and satellites. This configuration accurately replicates authentic satellite communication linkages.}

\subsection{Experimental Setup}

\subsubsection{Datasets}

We evaluate the accuracy of the inference system using the RSVQA dataset \cite{lobry2020rsvqa}. The RSVQA is a remote-sensing visual question-answering dataset. It is divided into low-resolution (RSVQA-LR) and high-resolution (RSVQA-HR). RSVQA-LR is one of the earliest VQA datasets for low-spectral-resolution remote sensing, comprising Sentinel-2 satellite images with a resolution of 256×256 pixels. These images are sourced from OpenStreetMap. RSVQA-HR, as a high-spectral-resolution dataset, consists of aerial RGB images captured by the USGS with a resolution of 512×512 pixels. 

We further utilize three classification datasets --- RESISC45 \cite{cheng2017remote}, AID \cite{Xia2017AID}, and WHU-RS19 \cite{Dai2011WHURS19} --- to evaluate the accuracy of the proposed framework. RESISC45 is a large-scale remote sensing image dataset comprising 45 distinct land use categories, with each category containing 700 images of size 256 $\times$ 256 pixels, resulting in a total of 31,500 images. The AID dataset is a large-scale aerial image collection that covers 30 scene categories and includes 10,000 remote sensing images. WHU-RS19 is another remote sensing image dataset, encompassing 19 scene categories, with approximately 50 images per category.

\subsubsection{Models}

We use the pre-trained Qwen2-VL 2B on the satellite and the pre-trained Qwen2-VL 7B on the ground station. \lizhrev{In industrial practice, an LVLM of approximately 2B size is deemed suitable for application on edge devices. For example, Google's Gemini Nano on Pixel 8 Pro with 1.8B and 3.25B parameters, respectively \cite{gemini}.} Qwen2-VL incorporates vision-language adapters to enhance efficiency and supports multi-image functionalities \cite{Qwen-VL}. These multi-modal inference capabilities provide robust support for our archive's implementation. For the image and instruction embedding models used in the dynamic archive, we employ GeoRSCLIP ViT-B-32 \cite{zhang2024rs5m} to generate embedding vectors. \lizhrev{All models used in experiments have not undergone any fine-tuning. Therefore, all Qwen2-VL models are general-purpose models and have not been specifically fine-tuned for remote sensing data.}

\subsubsection{Baselines}

\begin{figure*}[t]
    \centering
    \begin{subfigure}[b]{0.245\linewidth}
        \centering
        \includegraphics[width=\columnwidth]{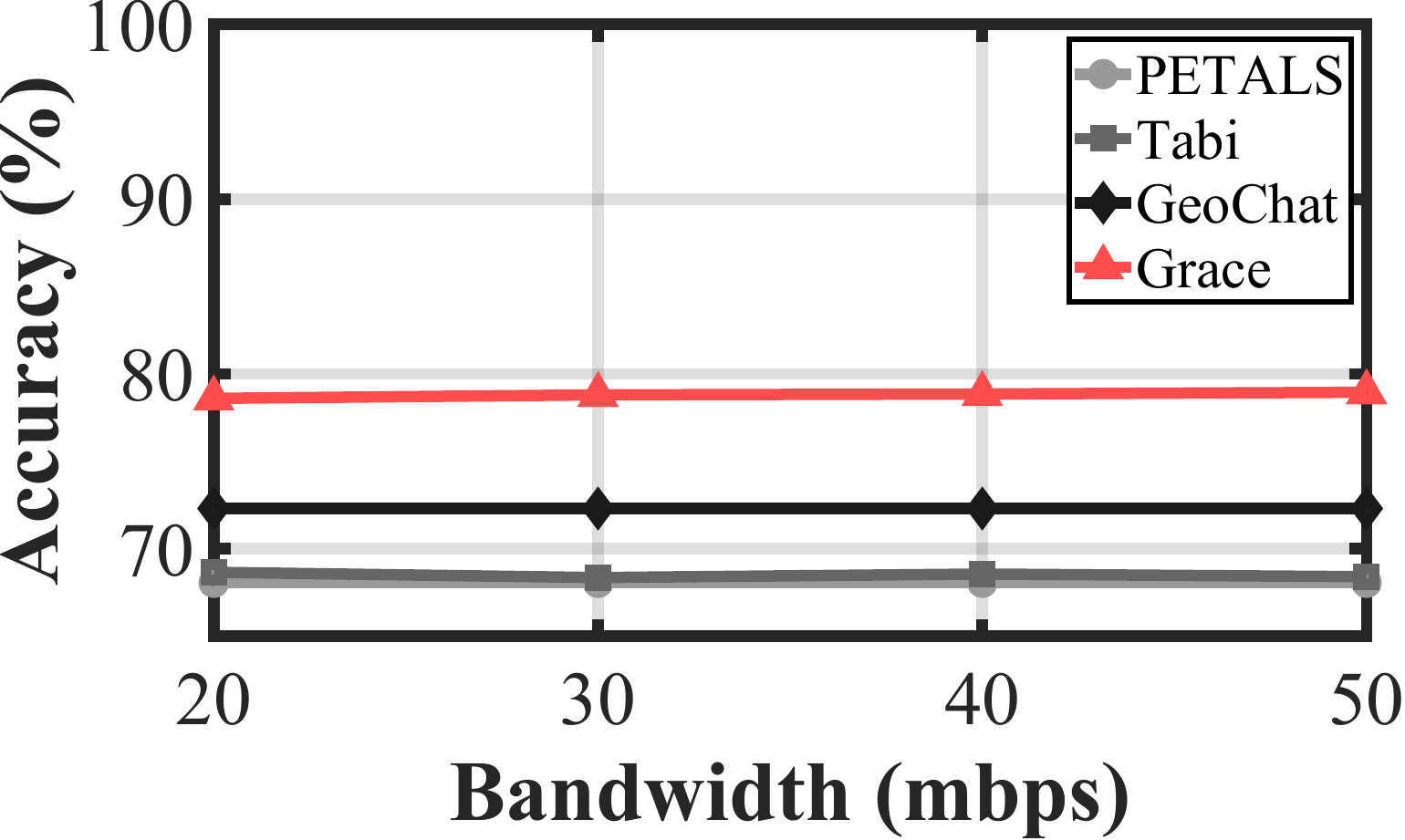}
        \caption{Test accuracy.}
        \label{fig:rate-acc}
    \end{subfigure}
    \hfill
    \begin{subfigure}[b]{0.245\linewidth}
        \centering
        \includegraphics[width=\columnwidth]{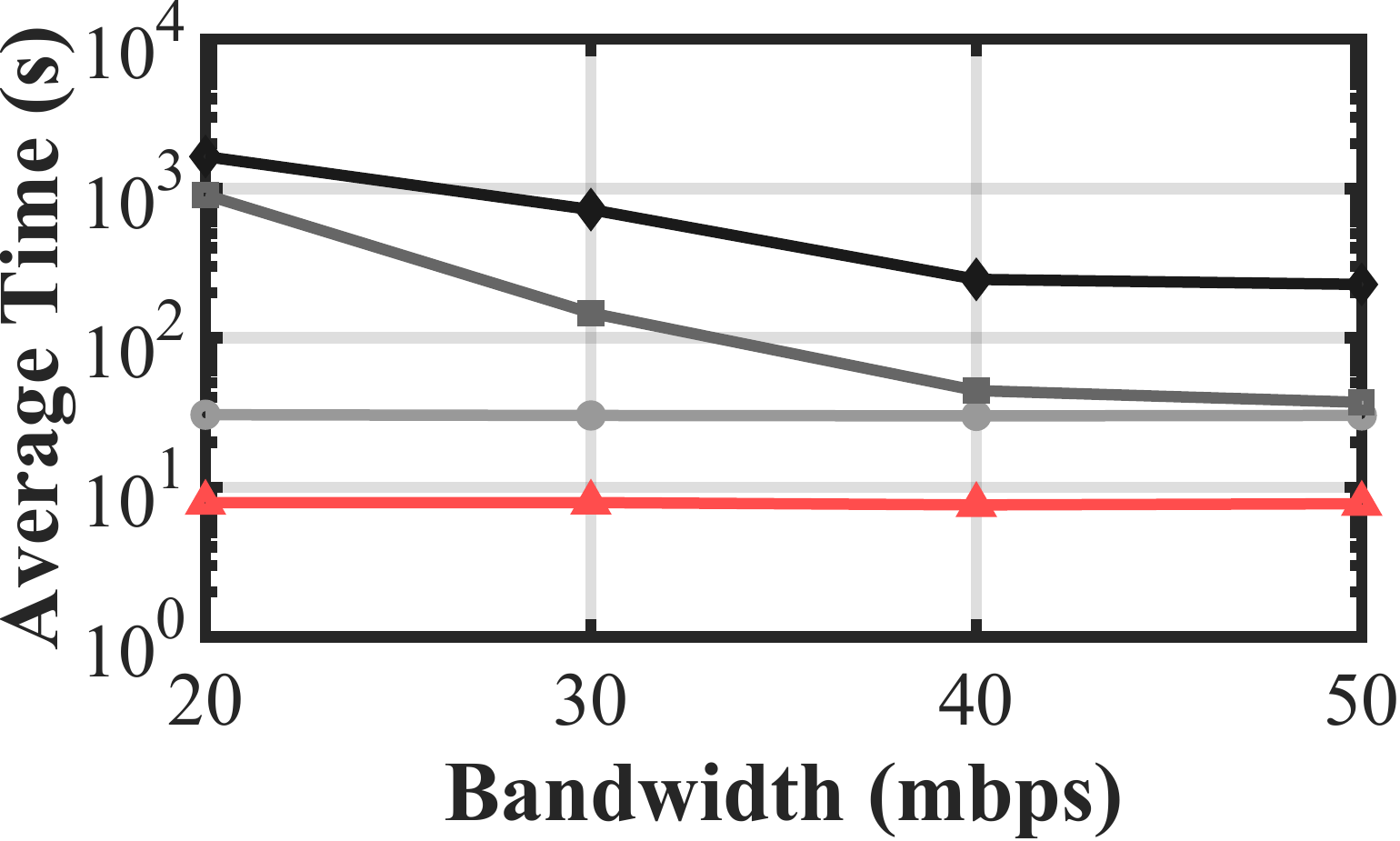}
        \caption{Average time.}
        \label{fig:rate-avg}
    \end{subfigure}
    \hfill
    \begin{subfigure}[b]{0.245\linewidth}
        \centering
        \includegraphics[width=\columnwidth]{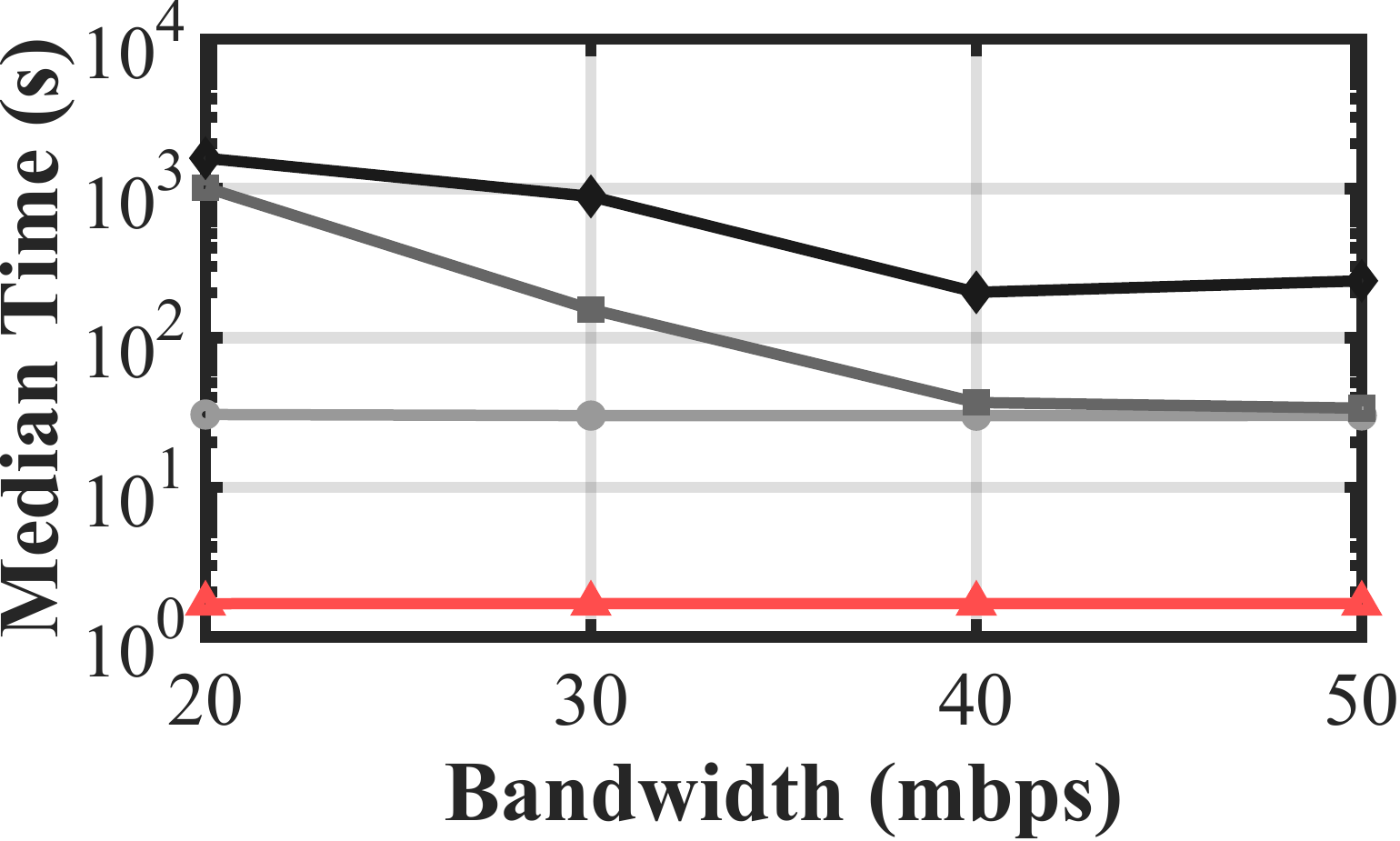}
        \caption{Median time.}
        \label{fig:rate-med}
    \end{subfigure}
    \hfill
    \begin{subfigure}[b]{0.245\linewidth}
        \centering
        \includegraphics[width=\columnwidth]{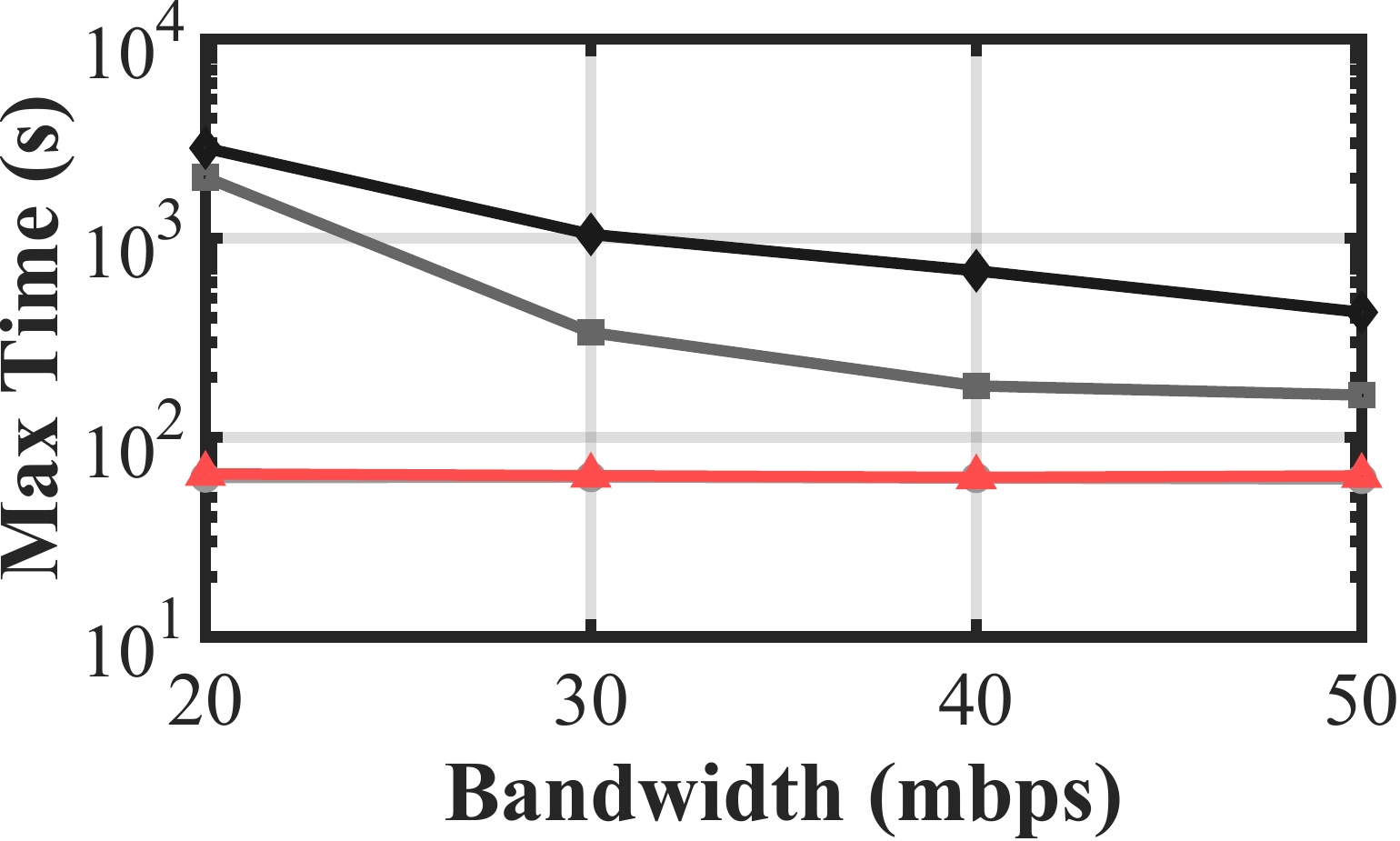}
        \caption{Max time.}
        \label{fig:rate-max}
    \end{subfigure}
    \caption{Test accuracy and inference time on RSVQA HR datasets under various bandwidth settings.}
    \label{fig:rate}
\end{figure*}

We compare Grace against several LVLM deployment baselines:

\begin{itemize}
    \item \textbf{PETALS} \cite{borzunov2022petals}: Borzunov et al. proposed PETALS, a system that joins multiple devices to collaboratively inference and fine-tune large language models. PETALS can run LLM reasoning on consumer-level GPUs.
    \item \textbf{Tabi} \cite{wang2023tabi}: Wang et al. propose a multi-layer cascaded efficient model serving system Tabi, which uses methods such as early return of simple queries, attention mechanism word pruning, and weighted multi-level ensemble learning.
    \item \lizhrev{\textbf{GeoChat} \cite{kuckreja2024geochat}: Kuckreja et al. get GeoChat by training a LLaVA v1.5 (7B) \cite{liu2023improvedllava} using a self-collected large-scale remote sensing dataset. The satellite only performs data collection, transmitting all data to the ground station via a wireless link. The ground station then conducts inference for each query using GeoChat.}
\end{itemize}

\subsubsection{Hyper-parameters}

We set the size of the relevant content $K$ to 5. In the matching test, the image threshold $T_M$ is set to 0.8, the instruction threshold $T_I$ to 0.94, and the minimum number of relevant records after filtering $T_K$ is set to 3. We establish a confidence threshold $T_{\mathrm{Conf}}$ of 0.75 for the cognitive test. We import all data in the training dataset into the ground archive. We initially utilize 20 random remote sensing images from the training set and their associated instruction-answer pairs for the satellite archive. Consequently, the satellite archive contains significantly less content than the ground archive and requires incremental updates during operation to meet query demands. During the experiment, the upper limit of the number of remote sensing images in the satellite archive is 20.

\section{Performance Evaluation}
\label{sec:performance-evaluation}

In this section, we evaluate the performance of Grace from three aspects: 1) Comparisons with four baselines to demonstrate the superiority of Grace. 2) Investigating the impact of network-related hyper-parameter on the performance. 3) Ablation study to show the necessity of each component in Grace, including dynamic knowledge archive, task dispatcher, and hierarchical transmission mechanism.

\subsection{Superiority of Grace}

\begin{figure*}[t]
    \centering
    \begin{subfigure}[b]{0.24\linewidth}
        \centering
        \includegraphics[width=\columnwidth]{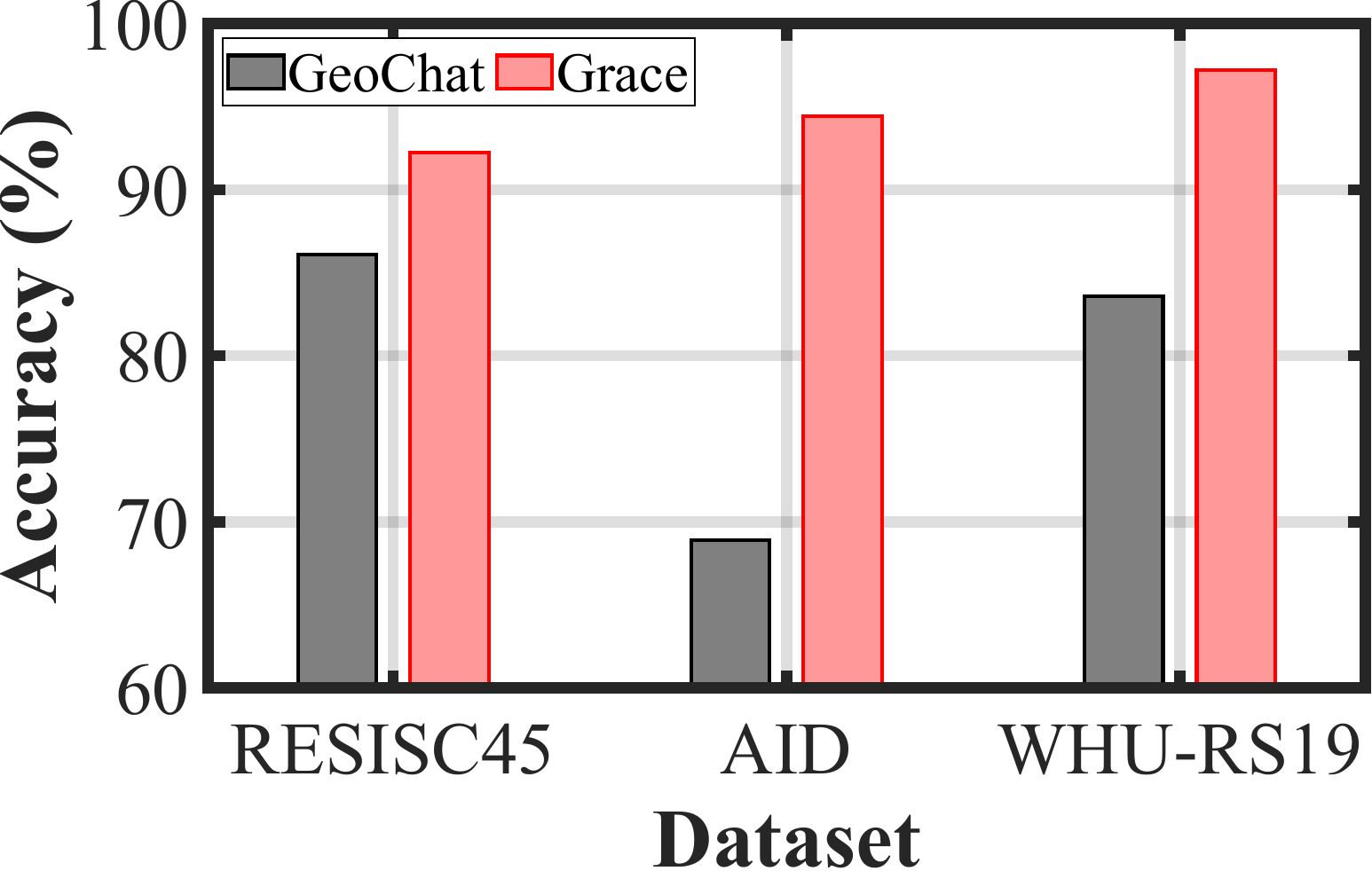}
        \caption{Fine-tuned LVLM.}
        \label{fig:geochat}
    \end{subfigure}
    \hfill
    \begin{subfigure}[b]{0.245\linewidth}
        \centering
        \includegraphics[width=\columnwidth]{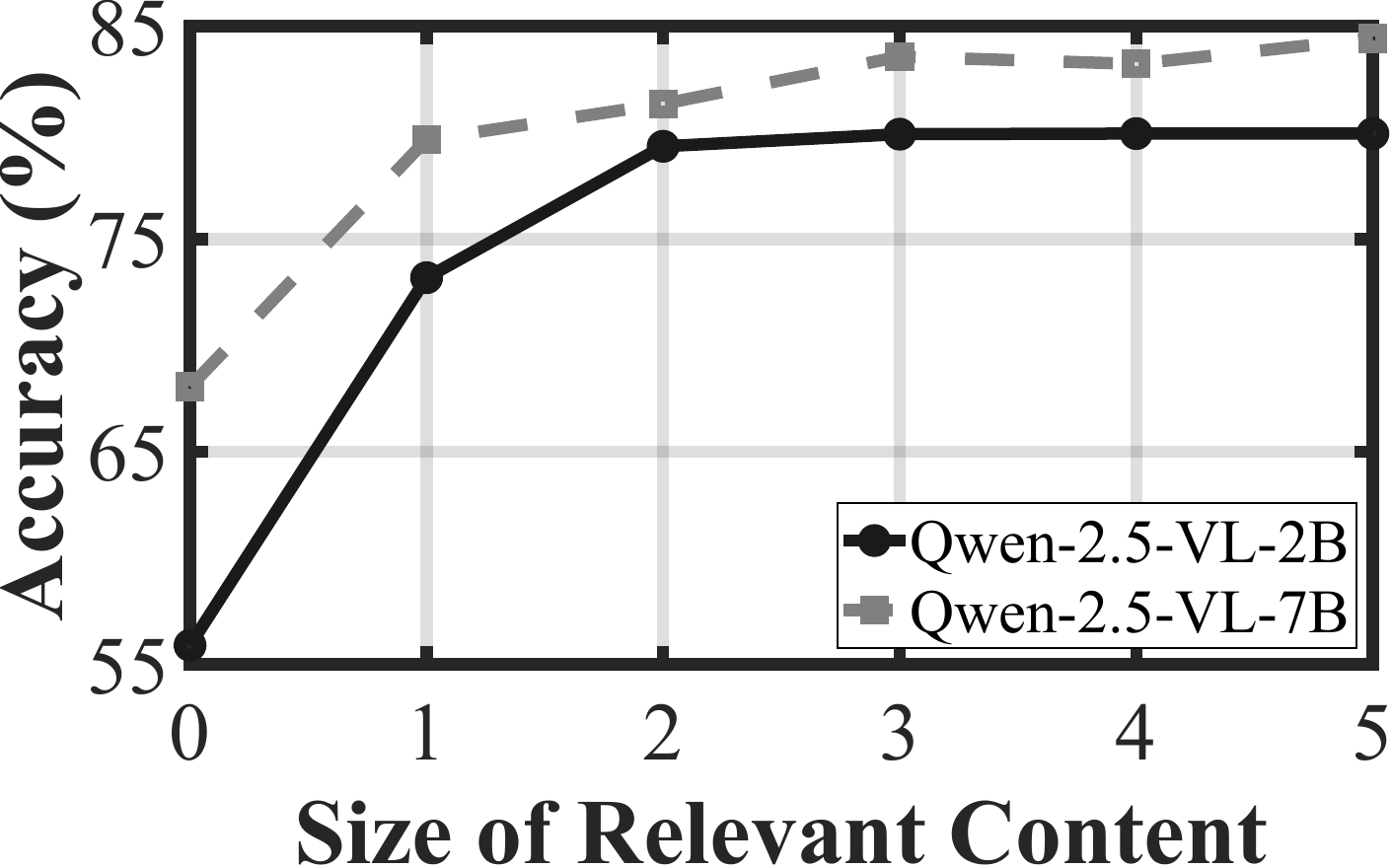}
        \caption{Relevant content.}
        \label{fig:absk}
    \end{subfigure}
    \hfill
    \begin{subfigure}[b]{0.245\linewidth}
        \centering
        \includegraphics[width=\columnwidth]{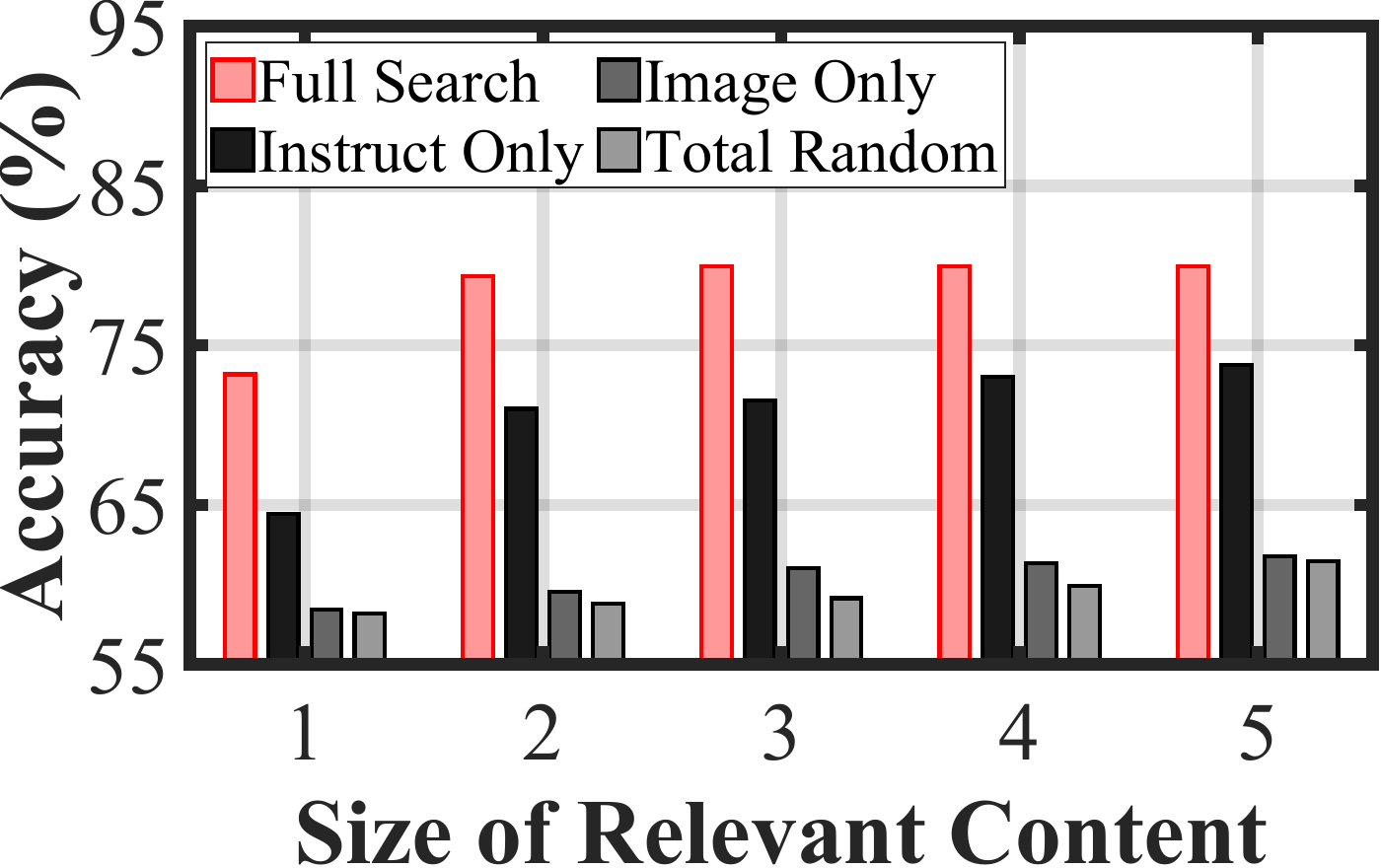}
        \caption{Qwen2-VL 2B.}
        \label{fig:abs2brand}
    \end{subfigure}
    \hfill
    \begin{subfigure}[b]{0.245\linewidth}
        \centering
        \includegraphics[width=\columnwidth]{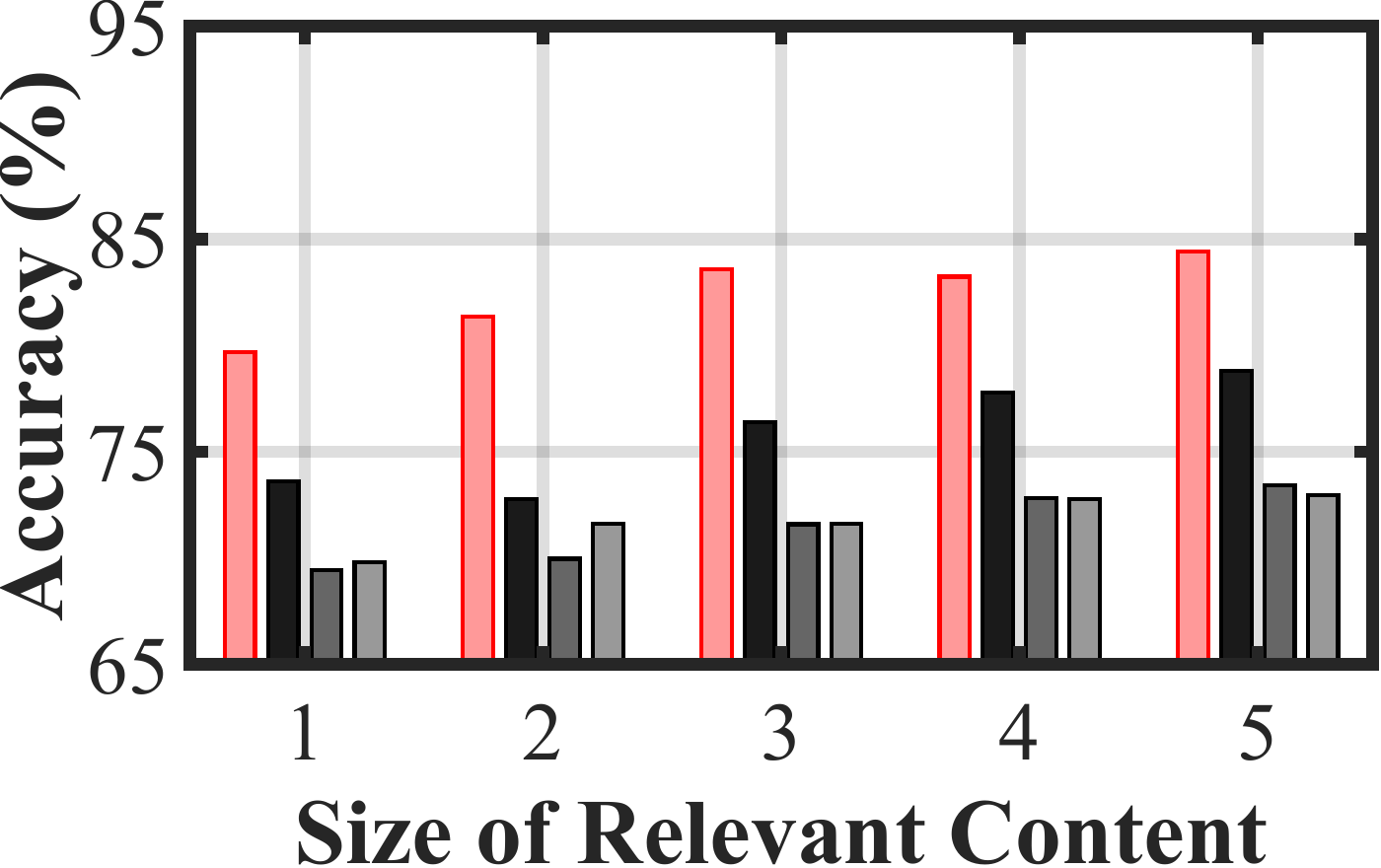}
        \caption{Qwen2-VL 7B.}
        \label{fig:abs7brand}
    \end{subfigure}
    \caption{\lizhrev{The test accuracy versus the fine-tuned LVLM, the size of relevant content ($K$), and different retrieval methods.}}
\end{figure*}


\lizhrev{As illustrated in Figure \ref{fig:main-acc}, we conduct a comprehensive evaluation of our proposed Grace framework against three representative baselines --- PETALS, Tabi, and GeoChat --- on both the Low-Resolution (LR) and High-Resolution (HR) RSVQA datasets. On the LR benchmark, Grace delivers an accuracy of 83.25\%, markedly higher than the sub-70\% performance of PETALS (67.25\%) and Tabi (67.07\%), and closely approaching the accuracy achieved by GeoChat, illustrating Grace’s competitiveness in scenarios where rapid, onboard reasoning is critical. In contrast, on the more challenging HR dataset, Grace attains an accuracy of 78.8\%, surpassing GeoChat’s 72.30\% by over 6 percentage points, and substantially outdoing both PETALS (68.05\%) and Tabi (68.35\%). This consistent superiority in the HR setting underscores Grace’s robustness when handling higher-fidelity imagery, a key requirement for precise remote sensing applications.}

We further analyze inference latency from task acquisition to result generation, with timing statistics summarized in Figures \ref{fig:main-avg}, \ref{fig:main-med}, and \ref{fig:main-max}. Three temporal metrics are examined: mean, median, and maximum processing times. \lizhrev{We appropriately reduce the orbital period of the LEO satellite as the limited number of test samples in datasets and the long time required for the satellite to complete one orbit. }

\lizhrev{Figure \ref{fig:main-avg} reveals that Grace achieves notable efficiency improvements in mean processing time compared to other baselines. 
Grace exhibits a mean latency of just 6.3s on LR and 7.9s on HR --- an impressive 76–95\% reduction relative to PETALS (30.3s / 30.2s) and GeoChat (30.7s / 722.3s), and an 88–95\% improvement over Tabi’s mean of 27.2s on LR and a prohibitive 146.7s on HR. These results demonstrate that Grace dramatically accelerates onboard inference, reducing reliance on intermittent ground links. 
Examining the median latencies in Figure \ref{fig:main-med} further highlights Grace’s stable performance: a median of 1.6s on LR and 1.6s on HR, compared to PETALS, Tabi, and GeoChat. The low median values indicate that for the vast majority of queries, Grace processes inputs almost instantaneously, with only a small fraction of outliers extending beyond this central tendency. 
The maximum latency analysis in Figure \ref{fig:main-max} provides critical insights into system robustness under worst-case scenarios. Grace exhibits constrained maximum delays of 61.1s (LR) and 64.2s (HR), markedly lower than the 336.4s (Tabi) and 1,045.8s (GeoChat) extremes observed in alternative approaches. }

These results highlight that Grace effectively balances enhanced accuracy with reduced inference times, making it a highly efficient solution for onboard inference in resource-constrained LEO satellite environments. By optimizing both accuracy and speed, Grace ensures reliable performance while minimizing the dependency on ground-based processing, thereby enhancing the overall efficacy of the satellite-ground collaborative inference framework.

\subsection{Micro-benchmarking}

\subsubsection{The Impact of Ground-Satellite Bandwidth.}



\lizhrev{Figure \ref{fig:rate} systematically demonstrates Grace's superior robustness to bandwidth fluctuations compared to conventional approaches. While ground-station-dependent GeoChat exhibit severe latency degradation under constrained network conditions --- evidenced by mean delays surpassing 1,635.3 seconds at 20 Mbps and only marginally improving to 228.2 seconds at 50 Mbps --- Grace maintains consistent performance, maintaining mean latency within a narrow 7.6–7.9 second range across all tested bandwidth scenarios. This constitutes a 200-fold performance improvement over previous ground-station workflows. Notably, the system achieves a median latency of 1.6 seconds without compromising accuracy, in contrast to bandwidth-agnostic baselines like PETALS that enforce fixed 30-second latency thresholds at the expense of significant precision degradation.

In worst-case operational scenarios, Grace's architectural innovations manifest distinct advantages: it limits maximum latency to 65.6 seconds under 20 Mbps constraints, outperforming GeoChat and Tabi by 43 and 31 times respectively. Significantly, while competing approaches demonstrate dramatic latency variability --- Tabi's maximum delays decrease by 71\% and GeoChat by 85\% with increased bandwidth --- Grace maintains stability within a 4\% fluctuation margin, establishing unprecedented consistency for real-time remote sensing applications. These experimental results validate Grace's unique capability to decouple system performance from network constraints, representing a critical advancement for operational systems deployed in heterogeneous connectivity environments.}

\subsubsection{Compare to Fine-tuned LVLMs.}

Current approaches to developing remote sensing-oriented LVLMs typically involve fine-tuning foundation models on domain-specific datasets to impart specialized geospatial knowledge. This section demonstrates that foundation models with RAG enables more generalization capabilities than pure fine-tuning approaches. We evaluate our method against GeoChat across three benchmark remote sensing datasets: RESISC45, AID, and WHU-RS19, with experimental results summarized in Figure \ref{fig:geochat}. \lizhrev{To ensure fair comparison, all models used in this experiment are evaluated in their original configurations without any fine-tuning.} Our approach achieves superior performance across all test scenarios, attaining 92.2\% accuracy on RESISC45 (+6.1\% improvement over GeoChat), 94.4\% on AID (+25.5\%), and 97.2\% on WHU-RS19 (+13.6\%). Notably, the performance gap widens significantly on complex scene understanding tasks – the 25.5\% accuracy gain on AID highlights our method's enhanced capability in handling fine-grained visual-semantic relationships.

This performance advantage stems from our external archive that dynamically integrates foundational visual understanding with on-demand knowledge retrieval, avoiding the catastrophic forgetting problem inherent in pure fine-tuning approaches. Unlike GeoChat's static parameterization of geospatial knowledge, our system adaptively retrieves relevant concepts from external knowledge bases during inference, enabling better handling of long-tail scenarios and novel objects unseen in training data.

\begin{figure*}[t]
    \centering
    \hfill
    \begin{subfigure}[b]{0.25\linewidth}
        \centering
        \includegraphics[width=\columnwidth]{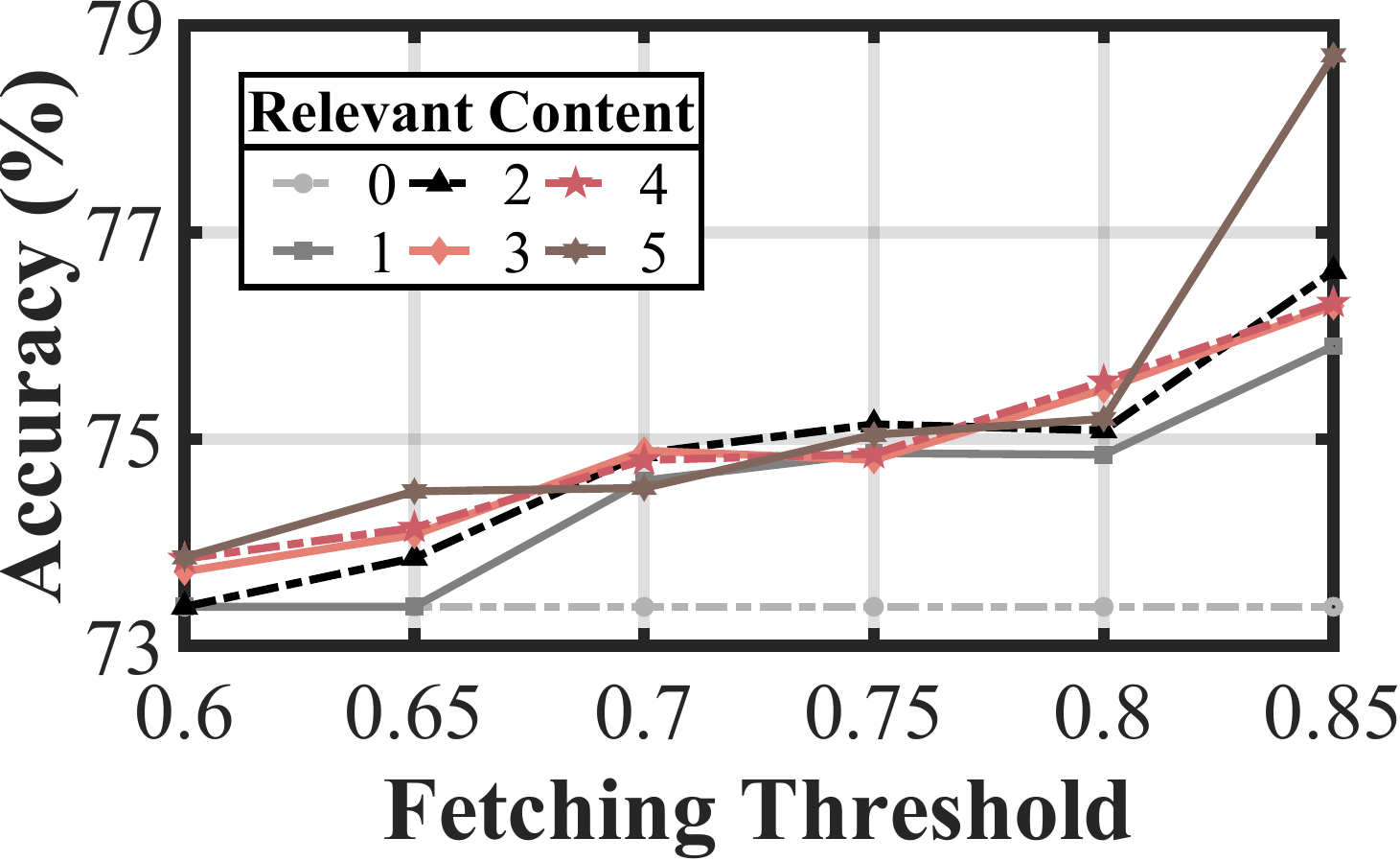}
        \caption{Image threshold.}
        \label{fig:abs2bimgthreacc}
    \end{subfigure}
    \hspace{0.6em}
    \begin{subfigure}[b]{0.25\linewidth}
        \centering
        \includegraphics[width=\columnwidth]{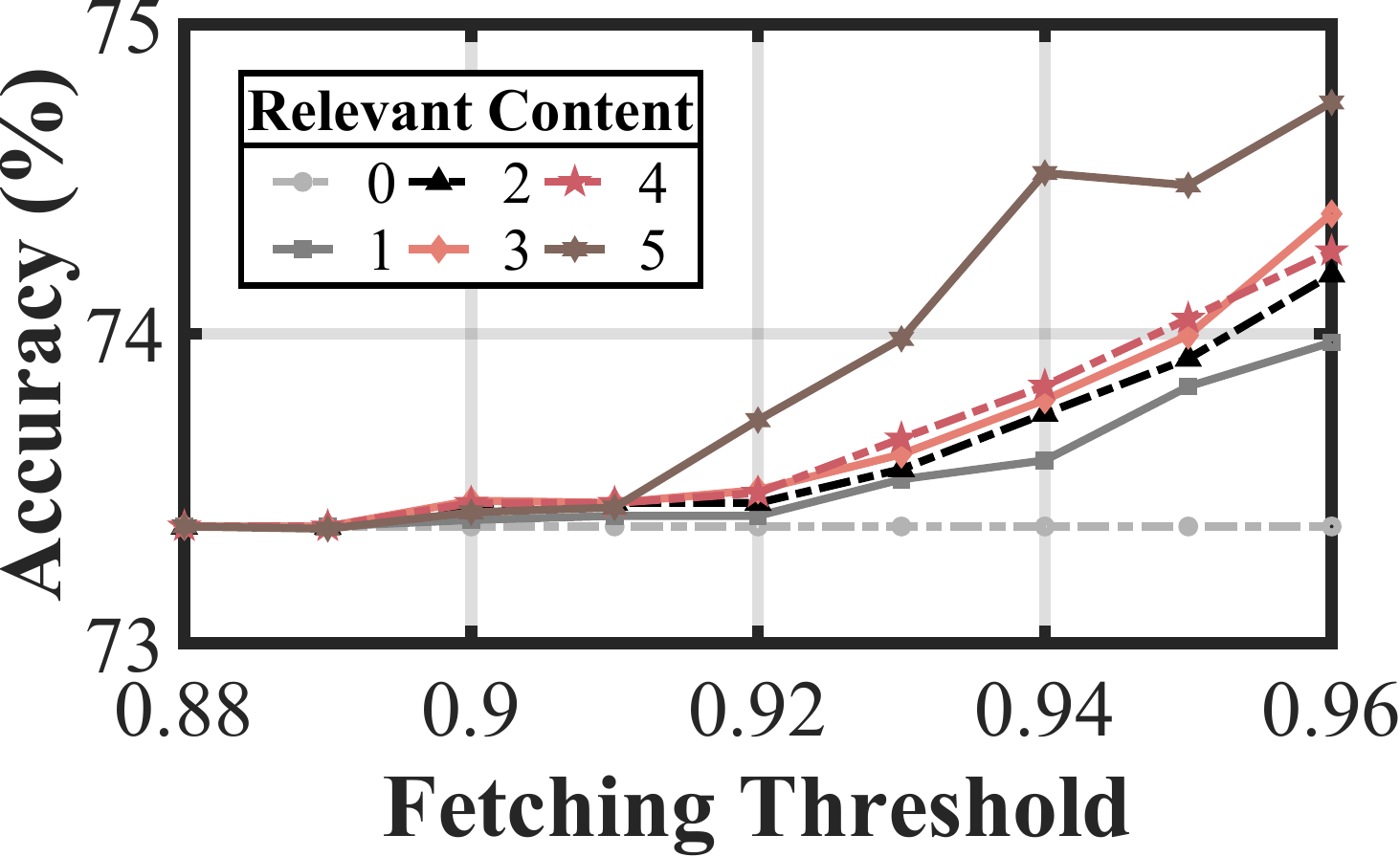}
        \caption{Text threshold.}
        \label{fig:abs2btxtthreacc}
    \end{subfigure}
    \hspace{0.6em}
    \begin{subfigure}[b]{0.25\linewidth}
        \centering
        \includegraphics[width=\columnwidth]{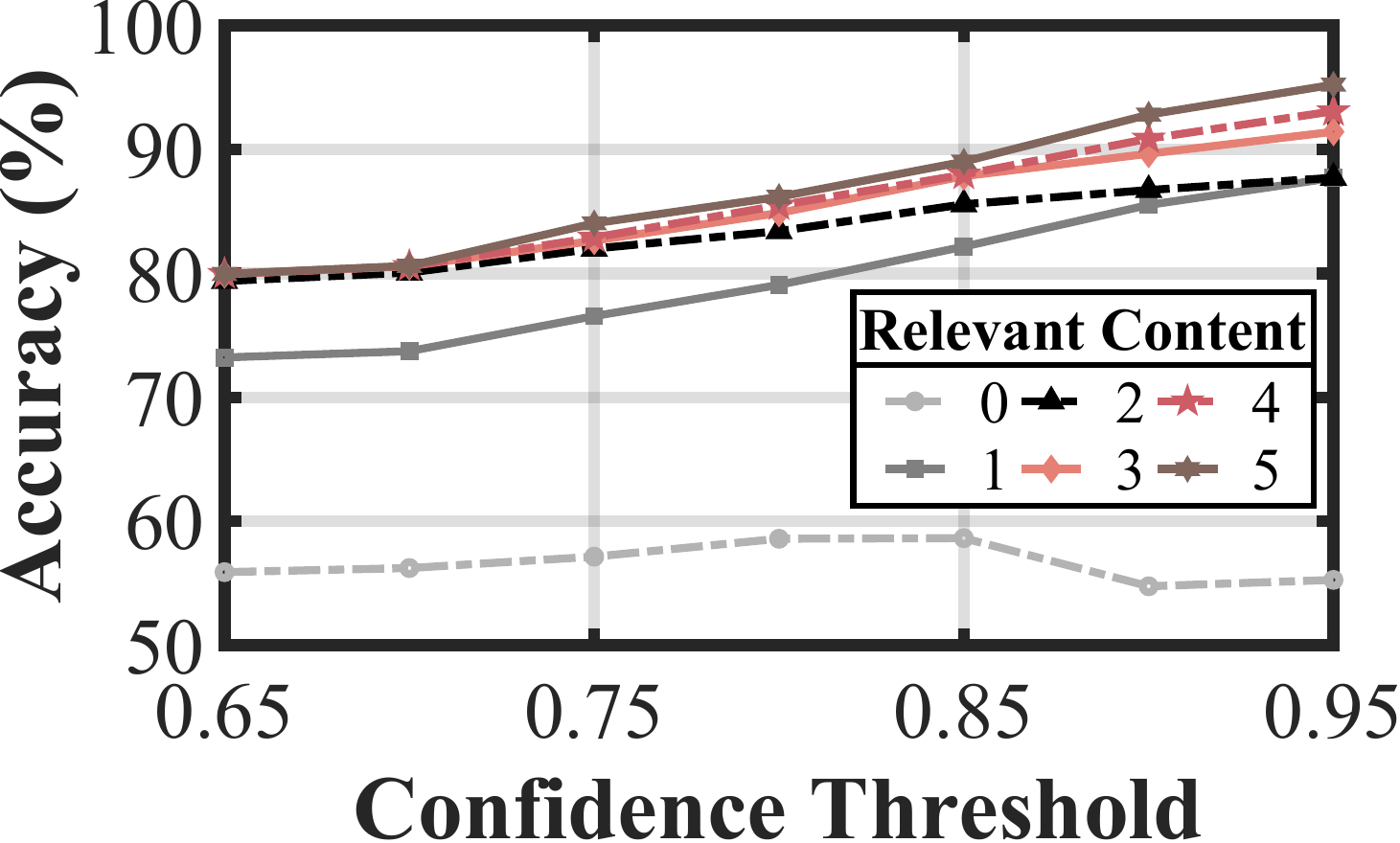}
        \caption{Confidence threshold.}
        \label{fig:abs2bconf}
    \end{subfigure}
    \hfill\null
    \caption{\lizhrev{The test accuracy versus the image (\(T_M\)) and the text (\(T_I\)) fetching threshold, and the confidence threshold (\(T_{\mathrm{Conf}}\)).}}
\end{figure*}

\subsection{Ablation Study}

\subsubsection{Impact of Size of Relevant Content.}

This section investigates the effect of using archives on the inference accuracy of LVLMs. We evaluated the inference accuracy of the LVLM without utilizing an archive corresponding to \(K=0\) relevant images. In this scenario, the LVLM relies solely on its internal knowledge to process queries. Additionally, we examined the impact of varying the number of relevant images from 1 to 5 on the inference process. Increasing \(K\) provides the LVLM with more examples, enhancing its ability to respond accurately to the instruction \(I_Q\). For comparison, we tested two models, Qwen2-VL 2B and Qwen2-VL 7B, both utilizing the complete RSVQA-LR training dataset as the archive.

The experimental results are depicted in Figure \ref{fig:absk}. Both the Qwen2-VL 2B and Qwen2-VL 7B models demonstrated a gradual increase in accuracy as \(K\) increased. This trend confirms that the archive effectively enhances the inference accuracy of LVLMs. Specifically, for the Qwen2-VL 2B model, the inference accuracy is 55.8\% without using an archive. Introducing a single relevant relevant image elevates the accuracy to 73.1\%, marking an improvement of 17.2\%. However, as \(K\) increased beyond one, the Qwen2-VL 2B model reached a plateau, maintaining a consistent accuracy of 79.9\% for \(K=4\) and \(K=5\).

The Qwen2-VL 7B starts with a higher inital accuracy of 68.05\% without the archive. Adding one relevant image boosts the accuracy by 11.6\% to 79.6\%. Unlike the 2B model, the 7B model continues to achieve performance gains even at \(K=5\), with a further 1.2\% increase compared to \(K=4\). Moreover, across all values of \(K\), the Qwen2-VL 7B model consistently outperformed the Qwen2-VL 2B model in inference accuracy. This indicates that the information provided by the archive alone is insufficient to bridge the inherent capability differences resulting from variations in model parameters. Consequently, it is necessary to complement the onboard satellite inference system with a ground-based inference system to achieve optimal performance.

\subsubsection{Impact of Retrieval Methods.}

This section explores how various retrieval strategies influence the inference accuracy of our collaborative inference framework for LEO satellites. We test four different retrieval methods:
\begin{itemize}
    \item \textbf{Full Search}: Utilizes our proposed multimodal retrieval algorithm, simultaneously querying both visual and textual modalities.
    \item \textbf{Instruct Only}: Retrieves solely based on text instructions, with the corresponding remote sensing images selected randomly.
    \item \textbf{Image Only}: Conducts retrieval exclusively on the visual modality, with text instructions chosen at random.
    \item \textbf{Total Random}: Selects relevant content entirely at random from the training dataset.
\end{itemize}
Figure \ref{fig:abs2brand} and \ref{fig:abs7brand} show results. The Qwen2-VL model demonstrates a significant performance improvement through multimodal fusion, with the 7B variant achieving 84.4\% accuracy under Full Search using five relevant images. This scaling advantage highlights the larger model’s enhanced capacity to integrate contextual cues. Multimodal retrieval proves critical across both scales, with Full Search outperforming Instruct Only and Image Only baselines by substantial margins. For the 7B model, the multimodal approach delivers an 11.0\% accuracy boost over image-based retrieval and 5.6\% over instruction-guided search, with similar trends observed in the 2B variant. The performance gap widens with added relevant images --- increasing from 5.3\% (1 record) to 10.8\% (5 records) between Full Search and Image Only for the 7B model, confirming progressive synergy between modalities.

These results substantiate the superiority of our multimodal retrieval algorithm. Notably, the image only approach does not offer advantage over the total random method. In contrast, combining instruction-based and image-based retrieval significantly enhances the accuracy of the LVLM’s inference. This demonstrates that our retrieval strategy is highly effective for the targeted inference scenarios, substantially improving the overall inference performance.

\subsubsection{Impact of Fetching Thresholds.}

In the matching test, our method filters out relevant records that show low cosine similarity with a given query. This section investigates how varying the similarity threshold affects inference accuracy. We conduct inference with the Qwen2-VL 2B model and record the image and instruction similarities of each relevant record for every query. We then define a range of thresholds \(T_M\) (from 0.60 to 0.85 in steps of 0.05) and retain only those relevant records whose similarity values exceed \(T_M\), provided the number of remaining records surpasses a lower bound \(T_K\).

Figure \ref{fig:abs2bimgthreacc} plots the resulting inference accuracy. Every curve with \(T_K > 0\) shows an ascending trend, indicating that higher thresholds \(T_M\) yield progressively better inference accuracy. The underlying rationale is that a larger cosine similarity value signals more substantial relevance, equipping the LVLM with more targeted information during inference. In contrast, when \(T_K=0\) (i.e., the matching test never filters out records), the accuracy curve forms a horizontal line below the other curves. This finding underscores the advantage of applying a matching test: it can eliminate specific queries unlikely to be inferred correctly, thereby conserving the satellite’s limited computing resources. Notably, a larger \(T_K\) typically correlates with higher accuracy.
We also perform instruction similarity tests under similar conditions. We vary the threshold \(T_I\) between 0.88 and 0.96. 
As shown in Figure \ref{fig:abs2btxtthreacc}, inference accuracy steadily increases with \(T_I\), mirroring the image similarity outcome. Thus, in both the visual and textual domains, there is a positive correlation between cosine similarity and inference accuracy: the closer the retrieved content is to the query, the greater the likelihood that the LVLM produces a correct answer.

\subsubsection{Impact of Confidence Threshold.}

Our cognitive evaluation framework establishes confidence scores as reliable indicators of LVLM inference validity. As shown in Figure \ref{fig:abs2bconf}, unaided inference ($K=0$) exhibits unstable accuracy trends across confidence thresholds, peaking at 58.6\% for \(T_{\mathrm{Conf}}=0.85\) before declining sharply. This contrasts with archival-supported scenarios ($K \geq 1$), where accuracy scales consistently with confidence thresholds --- rising from 73.2\% to 87.7\% for \(K=1\) as \(T_{\mathrm{Conf}}\) increases from 0.65 to 0.95. The 14.5\% improvement underscores the archive’s role in aligning high-confidence predictions with ground truth.  


These findings confirm that the cognitive test based on confidence scores effectively enhances inference accuracy within our Grace inference system, which leverages an external knowledge base. When using the archive, the positive correlation between confidence scores and accuracy underscores the importance of integrating confidence assessments with external knowledge sources to achieve reliable and accurate inferences in LEO satellite scenarios.

\subsubsection{Impact of Hierarchical Transmission.}

\begin{figure*}[t]
    \centering
    \begin{subfigure}[b]{0.245\linewidth}
        \centering
        \includegraphics[width=\columnwidth]{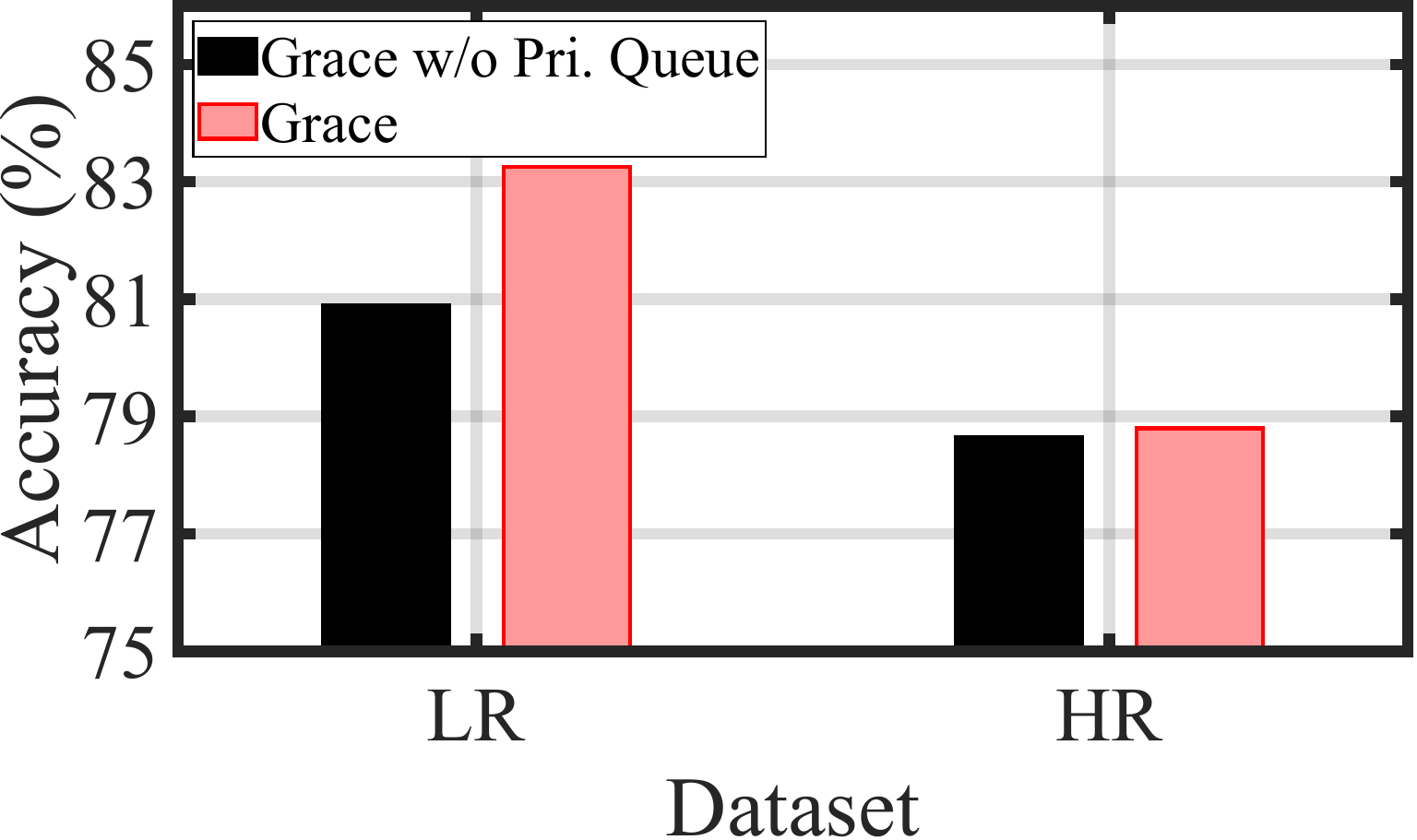}
        \caption{Test accuracy.}
        \label{fig:trans-acc}
    \end{subfigure}
    \hfill
    \begin{subfigure}[b]{0.245\linewidth}
        \centering
        \includegraphics[width=\columnwidth]{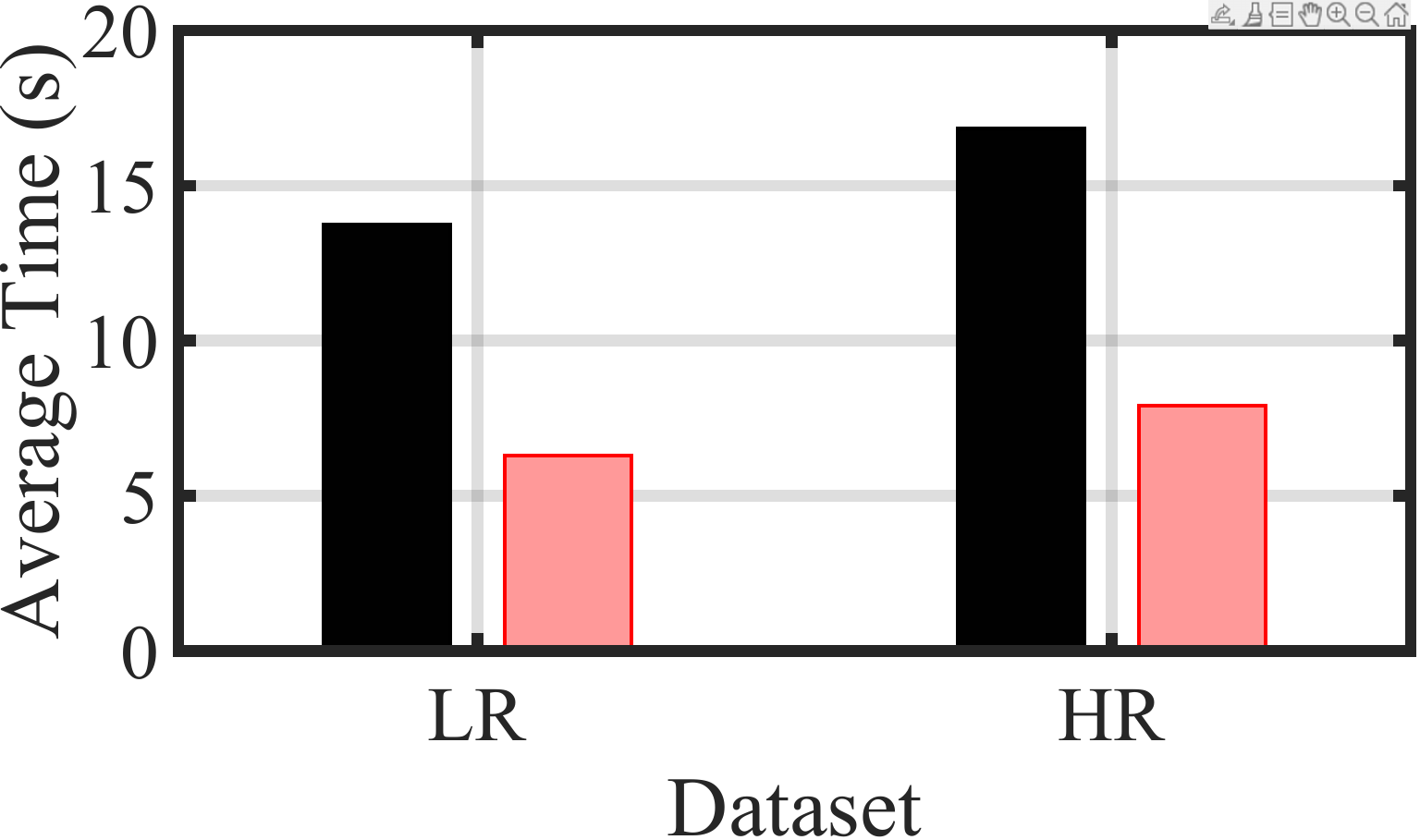}
        \caption{Average time.}
        \label{fig:trans-avg}
    \end{subfigure}
    \hfill
    \begin{subfigure}[b]{0.245\linewidth}
        \centering
        \includegraphics[width=\columnwidth]{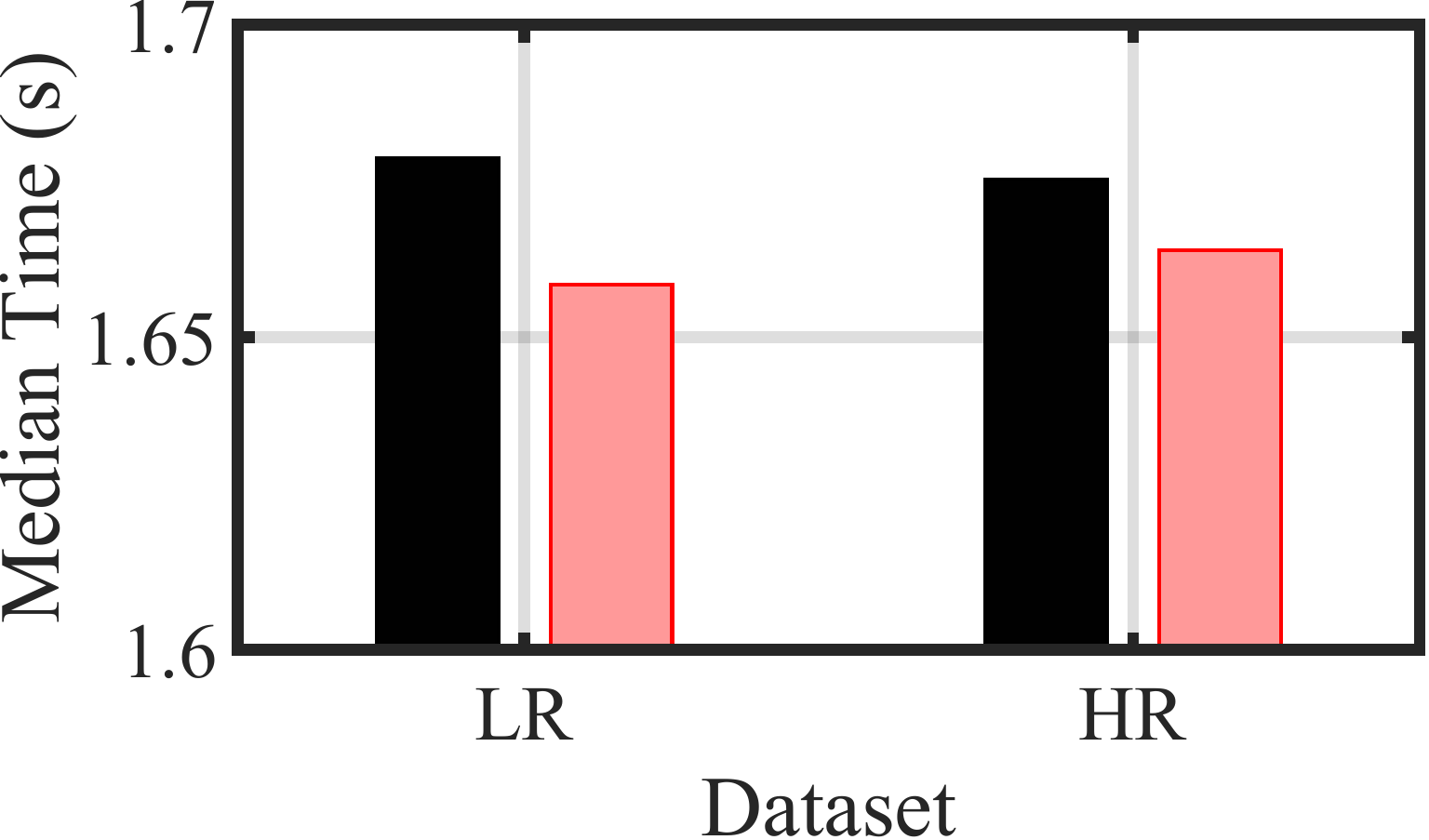}
        \caption{Median time.}
        \label{fig:trans-med}
    \end{subfigure}
    \hfill
    \begin{subfigure}[b]{0.245\linewidth}
        \centering
        \includegraphics[width=\columnwidth]{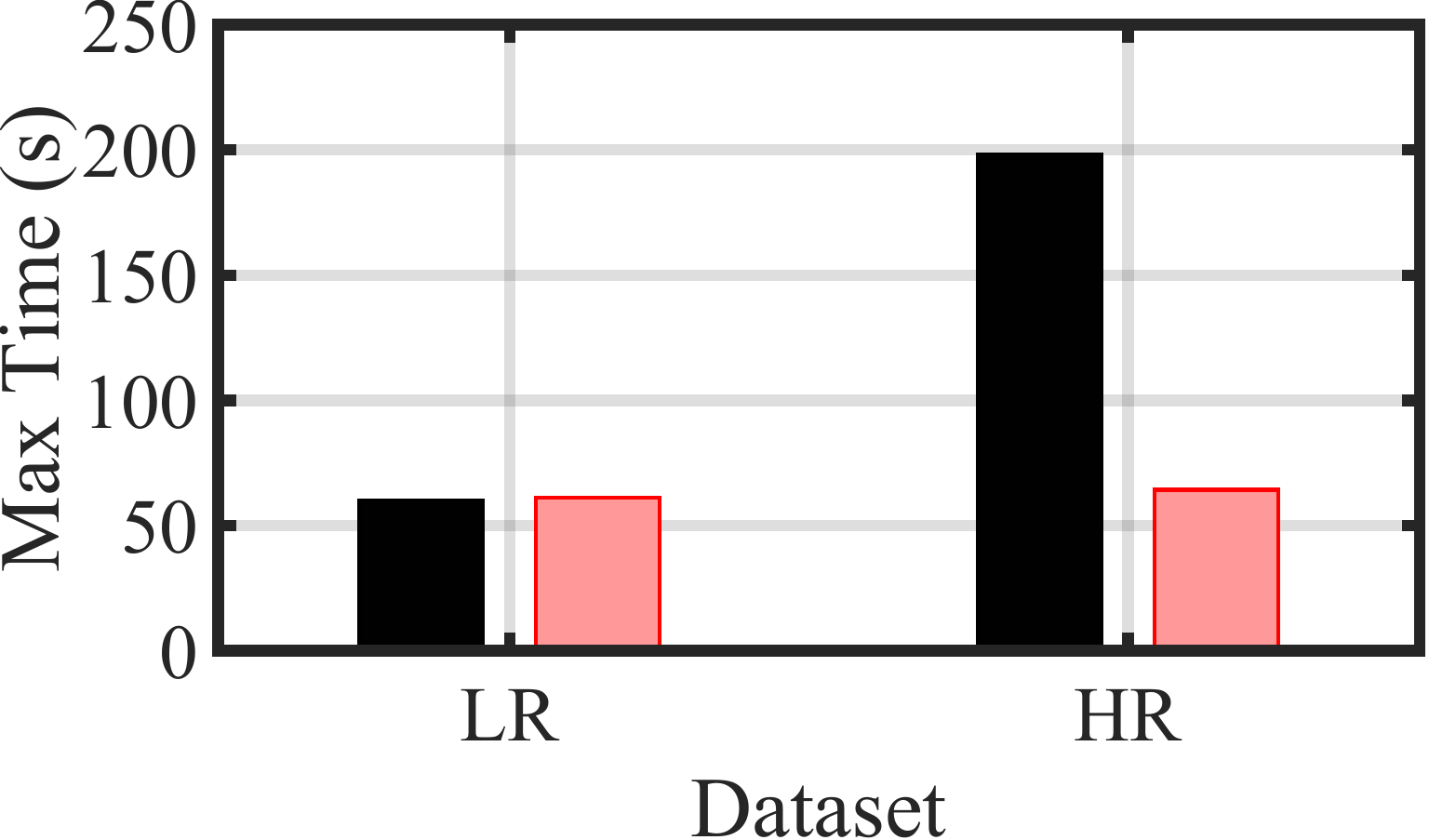}
        \caption{Max time.}
        \label{fig:trans-max}
    \end{subfigure}
    \caption{The impact of priority queue on test accuracy and latency on RSVQA LR and HR datasets. ``Pri. Queue'' indicates ``Priority Queue''.}
    \label{fig:trans}
\end{figure*}

In this section, we demonstrate the critical impact of priority queuing on optimizing temporal efficiency in both LR and HR processing pipelines. The Grace framework with priority queuing achieves a 53\% reduction in mean processing time compared to its non-prioritized variant, decreasing from 13.7s to 6.3s for LR data and from 16.8s to 7.9s for HR data, as shown in Figure \ref{fig:trans-avg}. This acceleration stems from intelligent task scheduling that prioritizes latency-sensitive operations while maintaining near-optimal median processing times.


\lizhrev{
The priority-aware approach further demonstrates robustness under peak workloads. As shown in Figure \ref{fig:trans-max}, HR processing exhibits a 67\% reduction in maximum latency (from 198.1s to 64.2s) through dynamic resource reallocation strategies that proactively mitigate computational bottlenecks. This constrained worst-case performance ($\leq$ 64.2s) is critical for ensuring predictable responsiveness in time-sensitive Earth observation applications, where both average-case efficiency and stability are essential for operational deployment. While median performance metrics remain comparable between configurations (Figure \ref{fig:trans-med}), the proposed method establishes a superior balance between throughput optimization and system reliability, making it particularly suitable for mission-critical remote sensing workflows.
}

\section{Related Work and Discussion}
\label{sec:related-work-and-discussion}

\subsection{LVLMs for Remote Sensing}

Transformer-based multimodal large models have received widespread attention in recent years for their strong generalization capabilities and ability to perform integrated inference with multimodal data \cite{liang2024survey}. These characteristics have motivated research into their application in remote sensing. Hu et al. proposed RSGPT \cite{hu2023rsgpt}, a vision-language model tailored for remote sensing. In addition, they constructed RSICap, a high-quality dataset for remote sensing image captioning. Zhang et al. introduced EarthGPT \cite{zhang2024earthgpt}, a multimodal large language model that leverages a vision-enhanced perception mechanism to integrate coarse-grained semantic and fine-grained detail information. They also released MMRS-1M, a large-scale dataset containing over one million image-text pairs. Meanwhile, Kuckreja et al. developed GeoChat \cite{kuckreja2024geochat}, a large vision-language model designed for remote sensing images, capable of multitask dialogue, especially region-level reasoning and visual localization in high-resolution remote sensing imagery.

Despite these advances, current studies on LVLMs in remote sensing primarily focus on training specialized large models using dedicated remote sensing datasets. This approach faces several drawbacks. First, training such models demands substantial hardware resources and lacks continuous update flexibility. Second, large LVLMs require high-performance computing to support inference, posing significant challenges for satellites operating as edge devices with constrained computational resources. In contrast, the Grace framework proposed in this paper features robust dynamic update capabilities, effectively balancing inference time and accuracy and offering a potential solution to some of these challenges.

\subsection{Collaborative Inference}

Collaborative inference leverages the robust computational power of the cloud alongside the low-latency benefits of edge computing, offering a potential solution to the challenges of poor edge inference performance on LEO satellites and high ground inference latency \cite{yao2022edge}. He et al. addressed issues such as data inefficiency, insufficient latency sensitivity, and inability to adapt to task load variations in existing cloud-edge collaborative inference systems \cite{he2024large}. They proposed an active inference-based method to optimize task offloading and resource allocation for large model inference tasks in cloud-edge systems. Wang et al. introduced Tabi \cite{wang2023tabi}, a multi-tier cascading efficient model service system that significantly reduces the latency and cost of cloud-edge collaborative inference without compromising accuracy. This is achieved through early exit simple queries, attention mechanism word pruning, and weighted multi-level ensemble learning. Borzunov et al. presented PETALS \cite{borzunov2022petals}, a framework supporting split inference operations across multiple computing nodes for large models.

Despite these advancements in collaborative inference in cloud-edge environments, practical deployment in LEO satellite networks remains challenging. Due to limited and unstable satellite-ground communication bandwidth, there is often no network connectivity between LEO satellites and ground stations for extended periods, severely restricting data transmission rates and system stability during the collaborative inference process. Designing a specialized collaborative inference framework tailored to the unique network conditions of LEO satellites is thus an urgent problem that needs addressing. This highlights the necessity for innovative approaches to overcome the inherent limitations of satellite-ground communications and ensure effective and reliable inference capabilities.

\subsection{Discussion}

\lizhrev{The proposed system does not incorporate an inter-satellite link (ISL) data transmission mechanism, primarily due to the absence of ISL communication capabilities in current Earth observation satellite constellations \cite{tao2024known}. While significant technological breakthroughs have been achieved in ISL applications for communications satellites (as exemplified by SpaceX's Starlink project \cite{starlinkplan}, which has demonstrated the feasibility of efficient data transmission between LEO satellites through laser-based ISLs), it is crucial to emphasize that these technological validations predominantly focus on communication satellite systems. The application scenarios and communication requirements of such systems differ from those of Earth observation satellites, which have yet to achieve large-scale deployment of ISL technology. Based on current technology maturity assessments, this study excludes ISL. Notably,  Grace’s modular framework enables future integration of mature ISL technologies through protocol upgrades and interface modifications, thereby potentially reducing satellite-to-ground communication latency.}

\section{Conclusion}
\label{sec:conclusion}

In this paper, we present Grace, a collaborative inference framework designed to address the unique challenges of remote sensing in LEO satellite systems. To the best of our knowledge, this is the first work to achieve near-realtime LVLM inference in LEO satellite networks. Our approach strategically balances resource constraints against the need for high-accuracy inference. We demonstrated that effective retrieval from a multimodal knowledge base capturing visual and textual information significantly improves inference accuracy on remote sensing image datasets. Matching tests filter out low-relevance records to conserve onboard resources, while cognitive tests leverage token-level probabilities to validate inference confidence. Experimental results show that increasing the number of relevant records and enforcing higher cosine similarity thresholds lead to marked improvements in inference performance, underscoring the critical importance of precise retrieval. Our dynamic update mechanism also ensures the satellite archive remains aligned with changing mission requirements, reducing unnecessary data transfers and enhancing overall system efficiency. Through these design considerations and empirical validations, our framework provides a robust, flexible, and resource-aware solution to enable accurate, highly efficient inference for LEO satellite applications.  As a potential future direction, we are looking forward to extending our \name to improve the performance of various applications such as distributed learning systems~\cite{lin2024efficient,hu2024accelerating,lin2024adaptsfl,zhang2025lcfed,lin2025hierarchical}, large language models~\cite{fang2024automated,lin2025hsplitlora,fang2025dynamic,duan2025leed,lin2024splitlora}, mobile edge computing~\cite{zhang2025robust,fang2024ic3m,tang2024merit,duan2025sample,lin2025hasfl}, etc, in LEO satellite networks.

\newpage
\bibliographystyle{ACM-Reference-Format}
\bibliography{sample-base}










\end{document}